# AIM: A User-friendly GUI Workflow program for Isotherm Fitting, Mixture Prediction, Isosteric Heat of Adsorption Estimation, and Breakthrough Simulation

Muhammad Hassan[a], Sunghyun Yoon[a], Yu Chen[a], Pilseok Kim[b], Hongryeol Yun[c], Hyuk Taek Kwon[d], Youn-Sang Bae[e], Chung-Yul Yoo[f], Dong-Yeun Koh[g], Chang-Seop Hong[c], Ki-Bong Lee[b], Yongchul G. Chung[a,h]*

[a] *School of Chemical Engineering, Pusan National University, Busan, 46241, Republic of Korea*

[b] *Department of Chemical and Biological Engineering, Korea University, 145 Anam-ro, Seongbuk-gu, Seoul 02841, Republic of Korea*

[c] *Department of Chemistry, Korea University, Seoul 02841, Republic of Korea*

[d] *Department of Chemical Engineering, Pukyong National University, 45 Yongso-ro, Nam-gu, Busan, 48513, Republic of Korea*

[e] *Department of Chemical and Biomolecular Engineering, Yonsei University, 50 Yonsei-ro, Seodaemun-gu, Seoul, 03722, Republic of Korea*

[f] *Department of Energy Systems Research and Chemistry, Ajou University, 206, World Cup-ro, Yeongtong-gu, Suwon 16499, Republic of Korea*

[g] *Department of Chemical and Biomolecular Engineering (BK-21 Plus), Korea Advanced Institute of Science and Technology (KAIST), Daejeon 34141, Republic of Korea*

*h Graduate School of Data Science, Pusan National University, Busan, 46241, Republic of Korea*

*Corresponding author. E-mail: drygchung@gmail.com (Y.G. Chung)

**Abstract**

Adsorption breakthrough modeling often requires complex software environments and scripting, limiting accessibility for many practitioners. We present AIM, a MATLAB-based graphical user interface (GUI) application that streamlines fixed-bed adsorption modeling and analysis through an integrated workflow, which includes isotherm fitting, enthalpy of adsorption estimation, mixture prediction, and multicomponent breakthrough simulations. AIM supports 13 isotherm models for isotherm fitting and includes Ideal Adsorbed Solution Theory (IAST) implementation (FastIAS') and extended Langmuir models for mixture isotherm predictions. Moreover, the isotherm models can be used to run non-isothermal breakthrough simulations along with isosteric enthalpies of adsorption from the Clausius-Clapeyron and Virial equations. Users can export detailed column and outlet profiles (e.g., composition, temperature) in multiple formats, enhancing reproducibility and data sharing among practitioners. We compared the breakthrough simulation results from AIM workflow and compared that with the experimental data in the literature for ternary gas mixture ($CO_2$/$H_2$/$N_2$) and found excellent agreement for outlet compositions and temperature profiles.

**Program summary**

Program title: AIM

Developer's repository link: https://github.com/Chung-Research-Group/AIM

Licensing provision: GPL-2.0

Programming language: MATLAB

Operating system: Windows, Mac, Linux

Nature of problem: Adsorption isotherm modelling; prediction of mixture gas adsorption and breakthrough curves for fixed bed system using ideal adsorption solution theory (IAST) or extended model mixture adsorption model

Solution method: Adsorption isotherm fitting; isosteric enthalpy of adsorption via Clausius-Clapeyron or Virial equation; IAST nonlinear equation solution; solution of dynamic fixed bed material and energy balances using finite volume method (FVM)

## 1. Introduction

Adsorption-based separation processes are considered an energy-efficient method for separating and purifying gases. These processes exploit the selective adsorption of one component over another on porous materials [1]. The industrial applications of adsorption processes include hydrogen gas purification, air separation, hydrocarbon separation or air dehumidification [2]. Adsorbent materials' performance can be evaluated using primary factors (e.g., textural properties, equilibrium isotherms). However, incorporating secondary factors (e.g., heat and mass transfer resistances, particle size, and pressure drop) is crucial for practical settings [3–7].

For fixed bed systems, breakthrough experiments can be conducted to evaluate the performance of adsorbent under dynamic conditions [8,9]. In a typical breakthrough experiment, a column packed with a bed of adsorbent is pressurized and purged with a carrier gas, which is usually inert. The gaseous mixture is then introduced at the column inlet, and the concentrations of the different components are recorded at the outlet. The adsorption of components in a gaseous mixture occurs as the mixture passes over the adsorbent. The adsorption process continues until the point when the adsorbent becomes saturated. Once the adsorbent becomes saturated, the adsorbates 'breakthrough' the column and the concentration profile of the adsorbate in the effluent as a function of time is termed as breakthrough curve. Breakthrough curves, which show how gas concentrations change over time at the outlet of an adsorbent column, are often used to characterize dynamic separation performance of adsorbent materials.

Breakthrough simulations have been used to quantify and rank the separation potential of various types of adsorbent materials for $CO_2$ capture [10,11], Xe/Kr separation [12,13], and hydrocarbon separations [14,15]. There have been several performance metrics developed based on the breakthrough results such as the breakthrough time of preferentially adsorbing gas [16] and separation potential [17]. The breakthrough time of the more adsorbing gas is the total duration after which the more adsorbing gas exits from the adsorber column. These metrics correlate well with the productivity of the fixed bed. The separation potential is the maximum amount of the less adsorbing gas which can be recovered from the adsorber column and is based on the mixture adsorption loadings at the fixed bed inlet conditions. The mixture adsorption loadings are typically estimated based on the Ideal Adsorbed Solution Theory (IAST) developed by Myers and Prausnitz [18]. The model assumes negligible axial dispersion and mass transfer resistance, which leads to sharper and nearly vertical breakthrough curves [17,19]. However, in many cases the condition of negligible axial dispersion is not achieved due to low gas velocities and increased molecular diffusivity [20]. Similarly, larger particle size and low values of intra-particle diffusivities can lead to non-negligible mass transfer resistance causing significant deviations from the adsorption equilibrium [21]. In such cases, a high-fidelity breakthrough simulation accounting for axial dispersion and mass transfer, and other adsorber parameters are necessary for accurate performance evaluation of adsorbent materials.

Several commercial simulation software packages are available for breakthrough simulations, such as Aspen Adsorption, gPROMS Adsorption Library, ProSim Dynamic Adsorption Column, and AVEVA's Process Simulation Adsorption Library. These software applications are often used in literature to understand and predict the dynamic separation behavior of adsorbent materials, but the lack of standardized computational workflow and transparency of implemented methods for isotherm fitting, mixture prediction, and availability of breakthrough simulation parameters hampers the reproducibility of simulation results [22]. Moreover, many of these simulations are carried out in commercial software packages which makes it difficult to share workflow and raw data with other researchers without the software license. To remedy the situation, there has been some effort in developing open-source software tools for adsorption applications, such as pyIAST [23] and pyGAPS [24] for isotherm fitting and mixture adsorption, RUPTURA [25] for isotherm fitting, mixture adsorption, and breakthrough simulation, pyAPEP [26], and ToPSAil [27] for isotherm fitting, mixture adsorption prediction, and pressure swing adsorption cycle simulation. However, these software applications require the preparation of programming script or text input file, which may limit accessibility for practitioners. Several Graphical User Interface (GUI) tools have also been developed to improve usability. For instance, IAST++ [28] supports isotherm fitting and mixture adsorption prediction, and IHoA [29] provides isosteric heat of adsorption prediction. However, these GUI tools also have some limitations: IAST++ does not support breakthrough simulation, and IHoA requires pre-fitted isotherm parameters and lacks built-in isotherm fitting, which may restrict usability.

To overcome this, we present a MATLAB-based (GUI) software package called 'AIM'. AIM provides an integrated workflow for adsorption isotherm fitting at single and multiple temperatures, isosteric enthalpy of adsorption estimation, mixture isotherm prediction, and multicomponent adsorption breakthrough simulations. Unlike existing tools such as pyIAST and RUPTURA, which require scripting expertise or access to the command-line interface, AIM offers a seamless, GUI-driven application for isotherm fitting, isosteric enthalpy of adsorption estimation, and breakthrough simulation. These features are divided into four modules called IsoFit, HeatFit, MixPred, and BreakLab (**Figure 1**). IsoFit and HeatFit modules are used for isotherm fitting for single and multiple temperatures, respectively. Both IsoFit and HeatFit modules incorporate 13 different isotherm models which includes Langmuir, Langmuir–Freundlich, Quadratic, and Brunauer–Emmett–Teller (BET), and others. The parameters estimated from these two modules can be used in MixPred module to calculate mixture adsorption loadings at specified pressures. These mixture predictions are based on the IAST and extended dual-site Langmuir (EDSL) models. The output files from IsoFit and HeatFit can be conveniently used to predict mixture adsorption using MixPred. Similarly, the parameters obtained from these modules can be readily imported into BreakLab module to simulate isothermal/non-isothermal breakthrough process for up to five components (1 non-adsorbing and 4 adsorbing) and to obtain temperature and concentration profiles in the column and at the outlet. The integrated workflow using AIM modules is shown in **Scheme 1**.

This article is structured as follows. First, we describe the file format and general guidelines for preparing IsoFit and HeatFit modules data input file. We then explain the different modules available in AIM along with the underlying models (**Section 3-6**). In **Section 7** we will provide the details of numerical methods employed for isotherm fitting, IAST equations solution, and breakthrough simulations. The results for different case studies conducted using AIM modules are provided in **Section 8** to demonstrate the performance of AIM. The codes and compiled GUI application software can be accessed from GitHub (https://github.com/Chung-Research-Group/AIM).

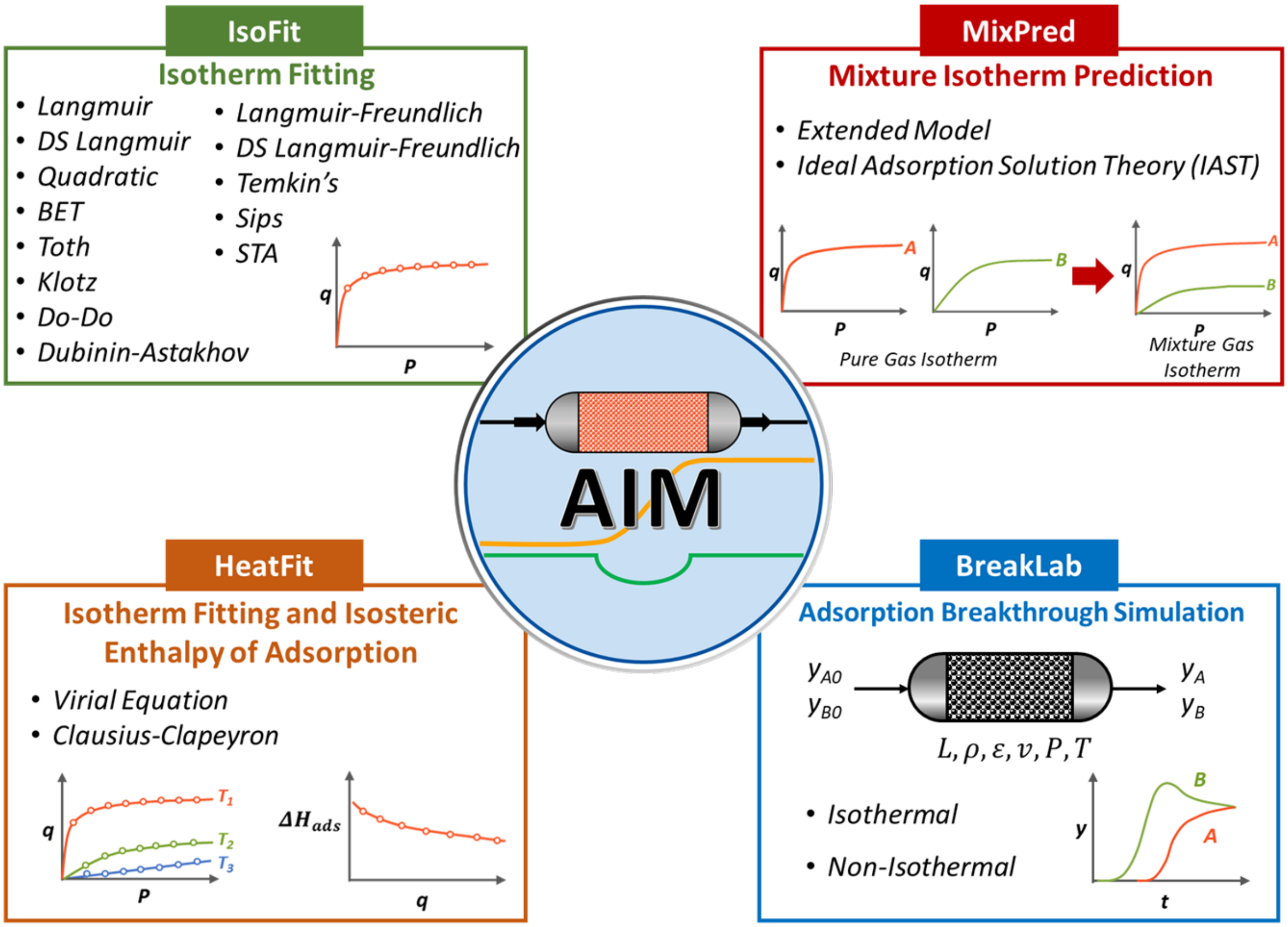

**Fig. 1.** Overview of AIM modules. **IsoFit** for isotherm fitting, **HeatFit** for multiple temperature isotherm fitting and isosteric enthalpy of adsorption prediction, **MixPred** for mixture isotherm predictions using the IAST and EDSL models, and **BreakLab** for multicomponent isothermal and non-isothermal breakthrough simulations

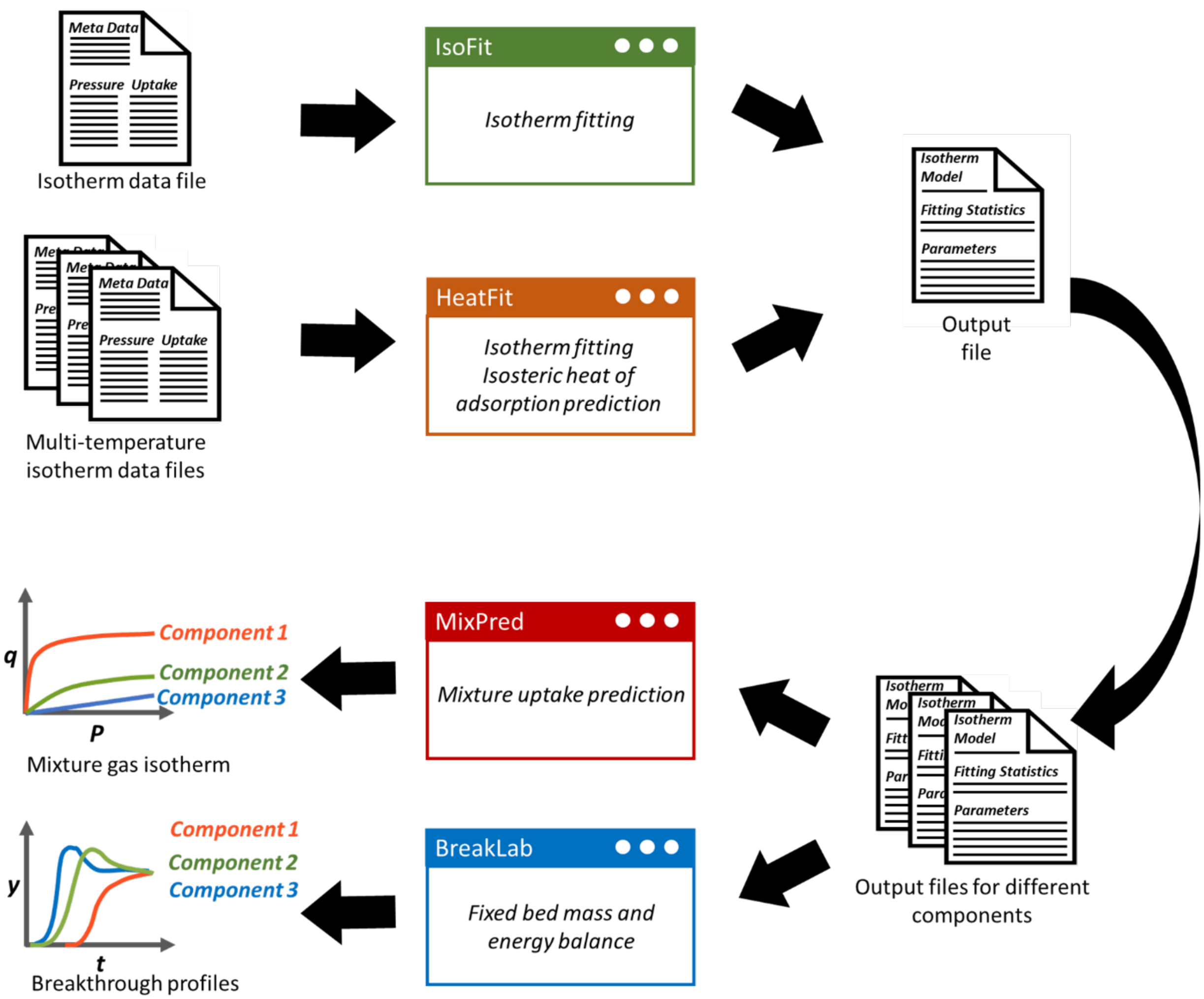

**Scheme 1.** Integrated workflow for isotherm fitting, mixture isotherm prediction and breakthrough simulation using AIM modules. IsoFit and HeatFit modules can read and fit the single and multiple temperature isotherm data for isotherm fitting, respectively. The fitting results from IsoFit and HeatFit can be saved in an output file, which contains the isotherm model, fitted parameters, and fitting statistics. The output file from HeatFit also contains the predicted isosteric enthalpy of adsorption value. The fitted isotherm model and the parameters can be directly loaded into MixPred and BreakLab modules by importing the output files for various components. MixPred mixture model uses the loaded isotherm model and parameters to calculate mixture adsorption isotherm. BreakLab uses the isotherm model and parameters to dynamically evaluate the mixture uptakes based on the fixed bed process conditions

## 2. Input and output data files

IsoFit and HeatFit modules read isotherm data from input file and generate output file containing isotherm fitting results. Both the input and output files have a well-defined and human-readable format to ensure accessibility to users. In this section, we outline the supported file formats, required keywords, and guidelines to facilitate isotherm data file preparation.

### 2.1. Input file description

IsoFit and HeatFit can import isotherm data from

- General input data files in the following formats: *.csv, *.txt, *.xlsx, *.dat.
- Adsorption Information Format (AIF) [30].

The general input data file contains tabular data with columns corresponding to pressure and adsorption uptake values. Additionally, the user can specify the meta information in the input file such as temperature value and units of pressure, adsorption uptake, and temperature. **Figure 2** shows a typical input file**.**

```
#units_pressure Pa
#units_loading mol/kg
#sat_pressure 2022
#temperature 298
#units_temperature K
72.4957     0.1400
161.1015    0.2451
229.5697    0.3326
318.1756    0.4201
394.6988    0.5427
503.4423    0.7177
579.9656    0.9103
```

**Fig. 2.** An example of a generic input data file for IsoFit and HeatFit modules. The header of the input data file contains meta information about the isotherm data. The meta information is followed by tabular data with two columns representing pressure and uptake values

We recommend following while preparing the generic input data file:

- The first column of the isotherm data contains pressure values, and the second column contains gas adsorption uptake (adsorption loading) values.

- The columns are separated either by a single tab, whitespace, or comma.
- The meta information is specified at the beginning of the file followed by isotherm data.
    - The meta information requires specific tags which are listed in **Table 1**. The module will only search for and parse these tags, and any other tags will be ignored.
- The tag names and values are specified in the following manner:

  Tag_1_Name  Tag_1_Value

  Tag_2_Name  Tag_2_Value

  …

In the case of importing data from a spreadsheet (*.xlsx), we recommend specifying the data in the first sheet. If the AIF file contains both adsorption and desorption data, only the adsorption data will be read. In case isotherm fitting using desorption data is required, we recommend using a separate input file. Finally, the temperature value is read only in HeatFit and not in IsoFit module.

In addition to isotherm data, IsoFit module also requires a manual input of saturation pressure ($P_0$) of the adsorbate for fitting the Dubinin-Astakhov, Klotz, and Do-Do isotherm models. The $P_0$ value is used to compute the relative pressure. The $P_0$ can be specified via one of the following methods:

- In the meta information of the isotherm data file using the tag listed in **Table 1**.
- In the app itself using the $P_0$ entry field.
- By directly specifying the relative pressure values instead of pressure values in the isotherm data file. In this case, specify $P_0 = 1$ in the $P_0$ entry field.

Moreover, in 'Auto' mode, the module attempts to fit all available isotherm models. For the three models that require $P_0$, we recommend verifying that a valid $P_0$ value is specified before relying on their fitting results.

**Table 1.** Tags name for specifying meta information in the input data files

| Description | General input data file | AIF |
|---|---|---|
| Pressure unit | #units_pressure | _units_pressure |
| Gas adsorption uptake unit | #units_loading | _units_loading |
| Saturation Pressure | #sat_pressure | _adsorp_p0 |
| Temperature value | #temperature | _exptl_temperature |
| Temperature unit | #units_temperature | _units_temperature |

### 2.2. Output file description

IsoFit and HeatFit generate an output file containing the isotherm model name, fitted parameters, and fitting statistics. Fitting statistics include the root mean square error (RMSE) and the coefficient of determinant ($R^2$) values. The output file from HeatFit also contains reference temperature used for fitting and fitted isosteric enthalpy of adsorption values. The output file uses custom extension '*.bliso' to facilitate integrated workflow within the modules. The fitted isotherm parameters in the output file can be directly loaded in MixPred and BreakLab modules for mixture adsorption prediction and breakthrough simulation, respectively. The user is advised not to modify the extension and keywords in the output file, as this can cause issues in isotherm parameter loading. **Figure 3** shows an example output file from the IsoFit module.

Particular attention must be given to the units of fitted parameters. The units of the fitted parameters depend on the units of the isotherm data. To ensure flexibility and general applicability across various datasets, IsoFit and HeatFit modules are designed to accommodate isotherm data in arbitrary user-defined units (e.g., Pa, kPa, or bar for pressure; mol/kg or mmol/g for gas uptake). To preserve this flexibility, the modules do not enforce a fixed set of units or perform automatic unit conversions, nor do the modules report fitted parameter units in the output file. Users must therefore verify unit consistency prior to result interpretation.

```
*************************************************************************
*********************IsoFit ISOTHERM FITTING RESULTS*********************
*************************************************************************

Isotherm Model: DS-Langmuir

RMSE:   0.004575
r^2:    1.000000

Parameters
q_sat_1     2.263656
b_1     1.811334e-05
q_sat_2     1.753061
b_2     5.744321e-07

*************************************************************************
```

**Fig. 3.** Typical output file from IsoFit, showing the fitted isotherm model, statistics, and parameters.

## 3. IsoFit – Isotherm fitting at single temperature

Adsorption isotherm describes the relationship between the bulk phase pressure and the equilibrium uptake of adsorbate molecules in adsorbent material at constant temperature. Adsorption isotherms can be from experimentally measured or obtained from running open-ensemble atomistic simulation, such as grand canonical Monte Carlo (GCMC) simulation. The data can be then fitted into different isotherm models which describe the mathematical relationship between adsorption uptake, pressure, and temperature. IsoFit module can load isotherm data, which is displayed in the tabular format and plotted for visualization (**Figure 4**). The user can choose different isotherm models for fitting the data. **Table 2** summarizes the isotherm model expressions and the respective fitting parameters. The module displays the fitted isotherm model curve and reports the fitted parameters in the output panel. The Isotherm Fitting Settings window allows users to define initial guesses, as well as lower and upper bounds for the fitting parameters. Additionally, the user can test multiple initial guesses using multistart feature to try different random initial guesses for isotherm fitting and select the best results.

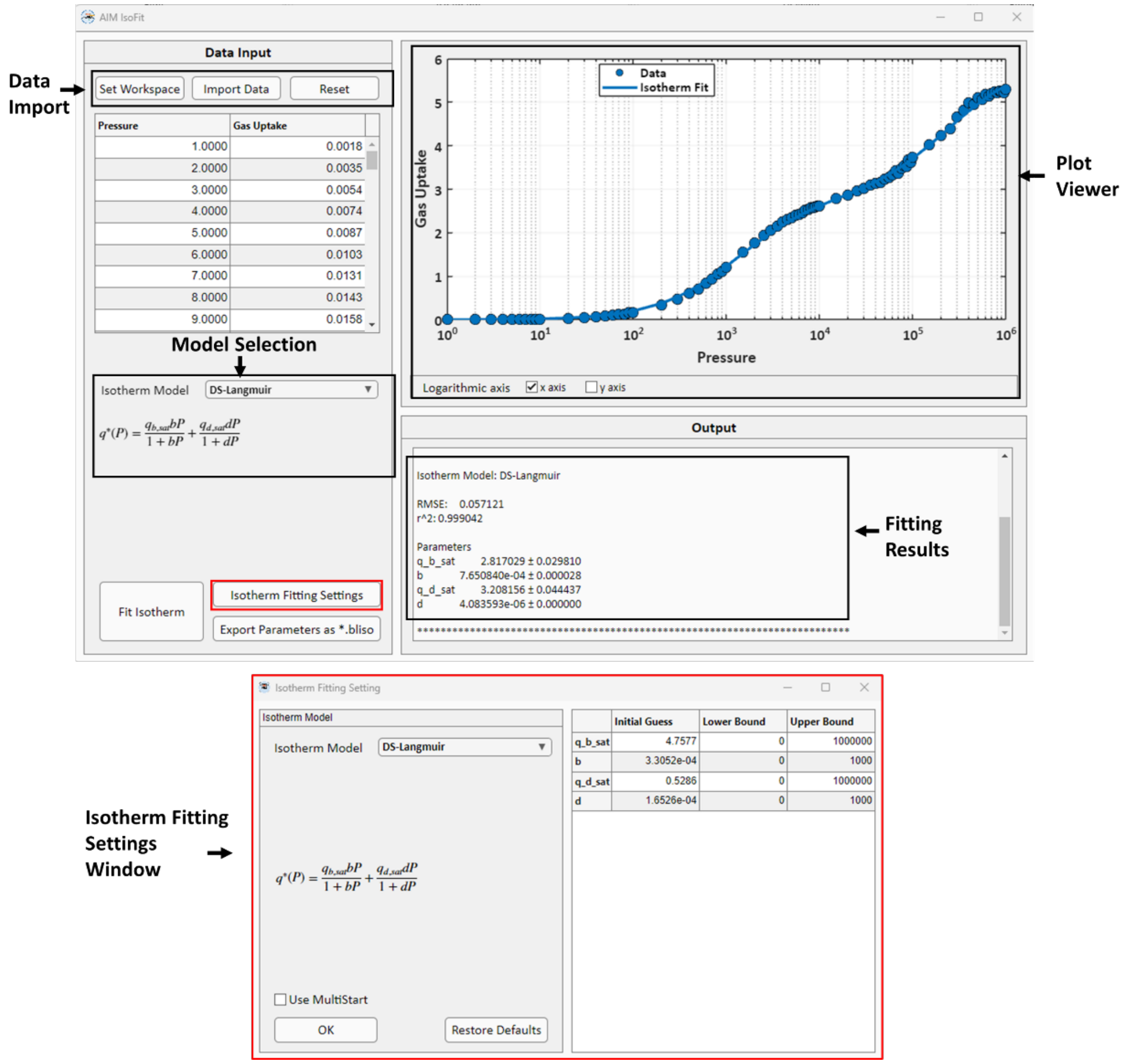

**Fig. 4.** IsoFit GUI with main isotherm fitting interface and Isotherm Fitting Settings dialogue box. **Data Input**: access to loading isotherm data and isotherm model selection for fitting. **Plot Viewer:** displays the fitted isotherm model curve and isotherm data. **Output:** numerical output of fitted parameters and the fitting statistics. **Isotherm Fitting Settings:** access to the initialization of the fitting procedure using user-defined initial guesses and upper and lower bounds for isotherm model parameters

**Table 2.** Isotherm models available in IsoFit

| Isotherm Models | Expression | Parameters |
|---|---|---|
| Langmuir | $q^* = \frac{q_{sat} bP}{1 + bP}$ | $q_{sat}, b$ |
| Dual-site Langmuir | $q^* = \frac{q_{sat,1} b_1 P}{1 + b_1 P} + \frac{q_{sat,2} b_2 P}{1 + b_2 P}$ | $q_{sat,1}, b_1,$ $q_{sat,2}, b_2$ |
| Langmuir-Freundlich | $q^* = \frac{q_{sat} bP^n}{1 + bP^n}$ | $q_{sat}, b, n$ |
| Dual-site Langmuir-Freundlich | $q^* = \frac{q_{sat,1} b_1 P^{n_1}}{1 + b_1 P^{n_1}} + \frac{q_{sat,2} b_2 P^{n_2}}{1 + b_2 P^{n_2}}$ | $q_{sat,1}, b_1, n_1$ $q_{sat,2}, b_2, n_2$ |
| Quadratic | $q^* = q_{sat} \left( \frac{bP + cP^2}{1 + bP + cP^2} \right)$ | $q_{sat}, b, c$ |
| Temkin | $q^* = q_{sat} \left( \frac{bP}{1 + bP} \right) + q_{sat} \theta \left( \frac{bP}{1 + bP} \right)^2 \left( \frac{bP}{1 + bP} - 1 \right)$ | $q_{sat}, b, \theta$ |
| BET | $q^* = \frac{q_{sat} bP}{(1 - cP)(1 - cP + bP)}$ | $q_{sat}, b, c$ |
| Sips | $q^* = \frac{q_{sat} (bP)^{\frac{1}{n}}}{1 + (bP)^{\frac{1}{n}}}$ | $q_{sat}, b, n$ |
| Toth | $q^* = \frac{qbP}{(1 + (bP)^n)^{\frac{1}{n}}}$ | $q_{sat}, b, n$ |
| Structural-Transition-Adsorption | $y(P) = \left( \frac{1 + b_{NP} P_{tr}}{1 + b_{NP} P} \right)^{q_{NP,sat}} \left( \frac{1 + b_{LP} P}{1 + b_{LP} P_{tr}} \right)^{q_{LP,sat}},$ $\sigma(P) = \frac{y^s}{1 + y^s},$ $q^* = (1 - \sigma) \left( \frac{q_{NP,sat} b_{NP} P}{1 + b_{NP} P} \right) + \sigma \left( \frac{q_{LP,sat} b_{LP} P}{1 + b_{LP} P} \right)$ | $q_{NP,sat}, b_{NP}$ $q_{LP,sat}, b_{LP}$ $s, P_{tr}$ |
| Dubinin-Astakhov | $x = \frac{P}{P_0}, \quad q^*(x) = q_{sat} e^{-\left( \frac{1}{K} \times \ln \frac{1}{x} \right)^n}$ | $q_{sat}, K, n$ |
| Klotz | $x = \frac{P}{P_0}, \quad s = Kx,$ $q^* = q_{sat} \frac{Cs\{1 - (1 + n)s^n + ns^{n+1}\}}{(1 - s)\{1 + (C - 1)s - Cs^{n+1}\}}$ | $q_{sat}, K, C, n$ |
| Do-Do | $x = \frac{P}{P_0},$ $q^* = q_{sat} (f \frac{K_1 x 1 - (1 + \beta) x^\beta + \beta x^{\beta+1}}{(1 - x)\{1 + (K_1 - 1)x - K_1 x^{\beta+1}\}}$ $+ (1 - f) \frac{K_2 x^\alpha}{1 + K_2 x^\alpha})$ | $q_{sat}, f, K_1$ $K_2, \alpha, \beta$ |

### 3.1. Langmuir isotherm model

The Langmuir isotherm model is based on the monolayer coverage of the adsorption sites with homogeneous interaction sites and first proposed by Irving Langmuir [31]. The Langmuir isotherm equation is:

$$q^* = \frac{q_{sat} bP}{1 + bP} \tag{1}$$

where $q_{sat}$ is the monolayer saturation capacity of adsorbent and $b$ is the Langmuir constant that describes the affinity of the adsorbate and adsorbent, and $P$ is the system pressure. The Langmuir isotherm given by equation (1) is only valid when the adsorbent surface is energetically homogenous.

### 3.2. Dual-site Langmuir isotherm model

Graham [32] proposed a dual-site Langmuir isotherm model to account for the heterogeneity of an adsorbent surface by incorporating two distinct adsorption binding sites. The dual-site Langmuir equation is:

$$q^* = \frac{q_{sat,1} b_1 P}{1 + b_1 P} + \frac{q_{sat,2} b_2 P}{1 + b_2 P} \tag{2}$$

where $q_{sat,1}, q_{sat,2}$ are the monolayer saturation capacities, and $b_1, b_2$ are the Langmuir constants corresponding to site 1 and 2, respectively.

### 3.3. Langmuir-Freundlich isotherm model

The Langmuir-Freundlich isotherm model was proposed by Sips [33] to account for the heterogenous nature of adsorbent surface. The model was later utilized to explain dissociative chemisorption of hydrocarbon on activated charcoal [34]. The Langmuir-Freundlich isotherm equation is:

$$q^* = \frac{q_{sat} bP^n}{1 + bP^n} \tag{3}$$

where $q_{sat}$ is the monolayer saturation capacity, $b$ is the Langmuir constant, and $n$ is the factor accounting for adsorbent surface heterogeneity. The model reduces to the Langmuir isotherm model when $n$ is 1.

### 3.4. Dual-site Langmuir-Freundlich isotherm model

The Langmuir-Freundlich isotherm model can be further modified to better represent a heterogenous adsorbent surface by incorporating an additional adsorption site. The resulting isotherm model is referred to as the dual-site Langmuir-Freundlich isotherm. The dual-site Langmuir-Freundlich isotherm equation is:

$$q^* = \frac{q_{sat,1} b_1 P^{n_1}}{1 + b_1 P^{n_1}} + \frac{q_{sat,2} b_2 P^{n_2}}{1 + b_2 P^{n_2}} \tag{4}$$

where $q_{sat,1}, q_{sat,2}$ are the monolayer saturation capacities; $b_1, b_2$ are the Langmuir constants, and $n_1$, $n_2$ are the surface heterogeneity factors corresponding to site 1 and 2, respectively.

### 3.5. Quadratic isotherm model

Statistical thermodynamics proposes that the general form of an isotherm model is the ratio of two polynomials of identical degrees [35].

$$q^* = q_{sat} \left( \frac{bP + 2cP^2 + 3dP^3 + \cdots}{1 + bP + cP^2 + dP^3 + \cdots} \right) \tag{5}$$

In the literature, the second order expansion of the above isotherm model is often employed, and referred to as quadratic isotherm model [36]:

$$q^* = q_{sat} \left( \frac{bP + 2cP^2}{1 + bP + cP^2} \right) \tag{6}$$

The quadratic isotherm model exhibits an inflection point: the loading is convex at low pressures (anti-Langmuir adsorption) but gradually changes concavity as the adsorbent saturates [37] (Langmuir adsorption), forming an S-shape.

### 3.6. Temkin isotherm model

Temkin adsorption isotherm model, similar to Langmuir model, assumes homogeneity of adsorption sites, but it also incorporates the adsorbate-adsorbate interactions based on the linear dependence of the heat of adsorption on coverage. IsoFit module incorporates the asymptotic approximation of the Temkin isotherm model derived by Simon *et al*. [38] based on the mean-field argument. The Temkin isotherm equation is:

$$q^* = q_{sat}\left(\frac{bP}{1+bP}\right) + q_{sat}\theta\left(\frac{bP}{1+bP}\right)^2\left(\frac{bP}{1+bP} - 1\right) \tag{7}$$

The parameter $\theta$ represents the strength of the adsorbate–adsorbate interactions. Values of $\theta < 0$ imply attraction, while $\theta > 0$ implies repulsion.

### 3.7. BET isotherm model

Brunauer, Emmett, and Teller (BET) isotherm model is a generalization of the Langmuir model to account for multilayer adsorption [39]. The BET model assumes that the adsorption sites are energetically homogenous, but the adsorbate molecules could form multiple layers without limit. The BET isotherm equation is:

$$q^* = \frac{q_{sat}bP}{(1-cP)(1-cP+bP)} \tag{8}$$

where $b$ and $c$ are the Langmuir constants for the first layer of adsorbate on the bare surface of the adsorbent and for the subsequent layers on top of the first layer, respectively.

### 3.8. Sips isotherm model

Sips isotherm model is a combination of the Langmuir and Freundlich isotherms and is expressed as [33,40]:

$$q^* = q_{sat} \frac{(bP)^{\frac{1}{n}}}{1 + (bP)^{\frac{1}{n}}} \tag{9}$$

where the parameter $n$ represents the adsorbent surface heterogeneity. When $n$ is 1, Sips model reduces to the Langmuir model, indicating that the adsorbent has homogenous binding sites. Both Sips (equation 9) and Langmuir-Freundlich (equation 3) isotherms are combinations of Langmuir and Freundlich models and are mathematically equivalent. We can recover equation 3 from equation 9 by setting $b_{LF} = (b_{Sips})^{1/n_{Sips}}$ , and $n_{LF} = 1/n_{Sips}$, where $b_{LF}$, $b_{Sips}$ and $n_{LF}, n_{Sips}$ are the Langmuir constants and surface heterogeneity constants for Langmuir-Freundlich and Sips models, respectively. Since both isotherm models are widely used in literature, we have implemented both models in IsoFit.

### 3.9. Toth isotherm model

Toth isotherm model is another empirical modification of Langmuir model [41,42], and is expressed as:

$$q^* = q_{sat} \frac{bP}{[1 + (bP)^n]^{\frac{1}{n}}} \tag{10}$$

where the parameter $n$ characterize the system heterogeneity and is usually less than 1. For the cases where $n = 1$, the model reduces to Langmuir model. The Toth equation satisfies both low-pressure and high-pressure limiting conditions. At low pressures the model reduces to Henry's law with Henry coefficient of $q_{sat}b$, while at high pressures the model yields a finite saturation limit equal to $q_{sat}$ .

### 3.10. Structural transition adsorption isotherm model

Flexible MOFs undergo adsorption-induced structural transition referred to as “gate-opening” or “breathing”, resulting in an S-shaped isotherms. Structural transition adsorption (STA) model is a statistically derived isotherm model that can be used to model adsorption isotherm of flexible adsorbent materials, such as in metal-organic framework (MOFs) [43]. STA equation has six parameters which are $q_{NP,sat}, b_{NP}, q_{LP,sat}, b_{LP}, s$, and $P_{tr}$. The STA equation is [43]:

$$q^*(P) = (1-\sigma)\left(\frac{q_{NP,sat} b_{NP} P}{1 + b_{NP} P}\right) + \sigma\left(\frac{q_{LP,sat} b_{LP} P}{1 + b_{LP} P}\right) \tag{11}$$

where subscripts $NP$ and $LP$ stand for narrow-pore and large-pore, representing the state of adsorbent before and after the adsorption-induced transition, respectively. $q_{NP,sat}, b_{NP}$, and $q_{LP,sat}, b_{LP}$ are the Langmuir isotherm model parameters for the $NP$ and $LP$ regions, respectively. $\sigma$ is an S-shaped function which smoothly connects the $NP$ and $LP$ isotherm curves and is expressed as,

$$\sigma(P) = \frac{y^s}{1 + y^s} \tag{12}$$

$$y(P) = \left(\frac{1 + b_{NP} P_{tr}}{1 + b_{NP} P}\right)^{q_{NP,sat}} \left(\frac{1 + b_{LP} P}{1 + b_{LP} P_{tr}}\right)^{q_{LP,sat}}, \tag{13}$$

where $s$ is the parameter representing the shape of adsorption-induced structural transition. A high value of $s$ indicate more abrupt and steeper transition between the $NP$ and $LP$ regions. $P_{tr}$ is the threshold or transition pressure, i.e. the pressure where the adsorbent structural transition occurs.

### 3.11. Dubinin-Astakhov isotherm model

Dubinin-Astakhov (DA) isotherm model has been applied to describe S-shaped (type V) adsorption isotherms which appear in flexible MOFs and water adsorption cases. DA model was developed based on the assumption that adsorption on microporous adsorbents is governed by a micropore filling [44]. The DA isotherm equation is [45]:

$$q^*(x) = q_{sat} e^{-\left(\frac{RT}{\epsilon} \times \ln\frac{1}{x}\right)^n} \tag{14}$$

$$x = \frac{P}{P_0} \tag{15}$$

where $P_0$ is the saturation pressure of the adsorbate, $q_{sat}$ is the saturation capacity of adsorbent when $P = P_0$, $R$ is the general gas constant, $T$ is the given temperature, $\epsilon$ is the characteristic free energy of adsorption, and $n$ is the parameter accounting for surface heterogeneity. IsoFit uses the following form of DA equation for isotherm fitting [46],

$$q^*(x) = q_{sat} e^{-\left(\frac{1}{K} \times \ln\frac{1}{x}\right)^n} \tag{16}$$

where $K$ is the lumped parameter representing the characteristic free energy of adsorption and temperature.

### 3.12. Klotz isotherm model

Klotz isotherm model could be also applied to fit an S-shaped isotherm. The model is based on the concept of Klotz originally developed for protein interaction with small molecules [47]. The Klotz isotherm model considers multi-layer adsorption process and assumes formation of adsorbate clusters which are in equilibrium with one another. The Klotz equation has four parameters which are $q_{sat}, K, C,$ and $n$. The Klotz equation is [48]:

$$q^*(s) = q_{sat} \frac{Cs\{1 - (1+n)s^n + ns^{n+1}\}}{(1-s)\{1 + (C-1)s - Cs^{n+1}\}} \tag{17}$$

$$s = Kx \tag{18}$$

$$x = P/P_0 \tag{19}$$

where $P_0$ is the saturation pressure of the adsorbate, $q_{sat}$ is the saturation capacity of adsorbent, $n$ is the maximal association number representing the degree of adsorbate cluster formation (i.e. number of adsorbate layers in cluster), $K$ is the clustering constant describing adsorbate-adsorbate interaction, and $C$ is the ratio of the affinity of the adsorbate and adsorbent and clustering constant $K$. A high value of $C$ indicates strong adsorbate-adsorbent interactions.

### 3.13. Do-Do isotherm model

Do-Do isotherm model mathematically describes the type IV isotherm, which is characterized by an initial formation of a few adsorption layers at lower pressures, followed by pore filling [49]. Do-Do model was originally developed for explaining adsorption and desorption of water on activated carbon [50]. The model assumes that the adsorption process initiates with formation and growth of water clusters via hydrogen bonding around the functional sites. These functional sites correspond to the location of functional groups. Micropore adsorption proceeds only after some critical size of the functional site bounded cluster is achieved [50,51]. Do-Do model has six parameters which are $q_{sat}, f, K_1, \alpha, \beta,$ and $K_2$. The Do-Do isotherm equation is [51]:

$$q^*(x) = q_{sat}\left(f\frac{K_1\sum_{n=1}^{\beta} nx^n}{1+K_1\sum_{n=1}^{\beta} x^n} + (1-f)\frac{K_2\sum_{n=\alpha+1}^{\beta} x^n}{K_2\sum_{n=\alpha+1}^{\beta} x^n + \sum_{n-\alpha=1}^{\beta} x^{n-\alpha}}\right) \tag{20}$$

$$x = P/P_0 \tag{21}$$

where $P_0$ is the saturation pressure of the adsorbate, $q_{sat}$ is the saturation capacity of adsorbent, $f$ is the fraction of functional sites, $K_1$ is the clustering constant at the functional site, $\beta$ is the maximal association number representing the degree of adsorbate cluster formation at the functional site, $\alpha$ is the critical size for micropore adsorption, and $K_2$ is the interaction constant between the water molecules in the micropore. In IsoFit, we use the following form of the model equation derived by Buttersack [48] using series summation given as:

$$q^*(x) = q_{sat}\left(f\frac{K_1x\{1-(1+\beta)x^{\beta}+\beta x^{\beta+1}\}}{(1-x)\{1+(K_1-1)x-K_1x^{\beta+1}\}} + (1-f)\frac{K_2x^{\alpha}}{1+K_2x^{\alpha}}\right);\ \beta > \alpha \tag{22}$$

$$x = P/P_0 \tag{23}$$

**4. HeatFit – Isotherm fitting at multiple temperature**

In the previous section, we discussed the isotherm models at a single temperature. The adsorption loadings at different temperatures are related to the isosteric enthalpy of adsorption denoted as $\Delta H_{ads}$, which is a specific case of the enthalpy of adsorption that is determined at constant adsorption loading (i.e., fixed coverage). This value can be useful for modeling non-isothermal system. The isosteric enthalpy of adsorption can be obtained by fitting multiple temperature isotherm data for the isotherm model via the Clausius-Clapeyron equation and Virial equation.

The isosteric enthalpy of adsorption as a function of coverage can be useful in understanding how strongly molecules are bound to different coverages. Detailed discussion of the isosteric enthalpy of adsorption can be found in the literature [52,53]. It is worth mentioning that a closely related quantity called isosteric heat of adsorption, generally represented by $Q_{st}$, is also often used in the adsorption literature. The relationship between isosteric enthalpy of adsorption ($\Delta H_{ads}$) and isosteric heat of adsorption $Q_{st}$ is:

$$Q_{st} = -\Delta H_{ads} \tag{24}$$

HeatFit module can be used to calculate the isosteric enthalpy of adsorption based on user-provided isotherm data (**Figure 5**). The module supports two different models, Clausius-Clapeyron and Virial, to obtain the isosteric enthalpy of adsorption. To run the module, the user needs to first load multi-temperature isotherm data and the corresponding temperature values. The loaded data will be displayed in the tabular format and plotted in the GUI. When the user pushes 'Fit Isotherm' button, the isotherm fitting procedure will be carried out for the chosen isotherm model. The users can define initial guesses, as well as lower and upper bounds for the fitting parameters in the Isotherm Fitting Setting window. Additionally, the user can test multiple initial guesses using multistart features, which enables HeatFit to generate different random initial guesses for isotherm fitting and select the best results. Once the isotherm fitting is complete, the module plots the fitted isotherm model curve and reports the fitted parameters in the Output panel. The resulting isotherm parameters can be saved in several different file formats.

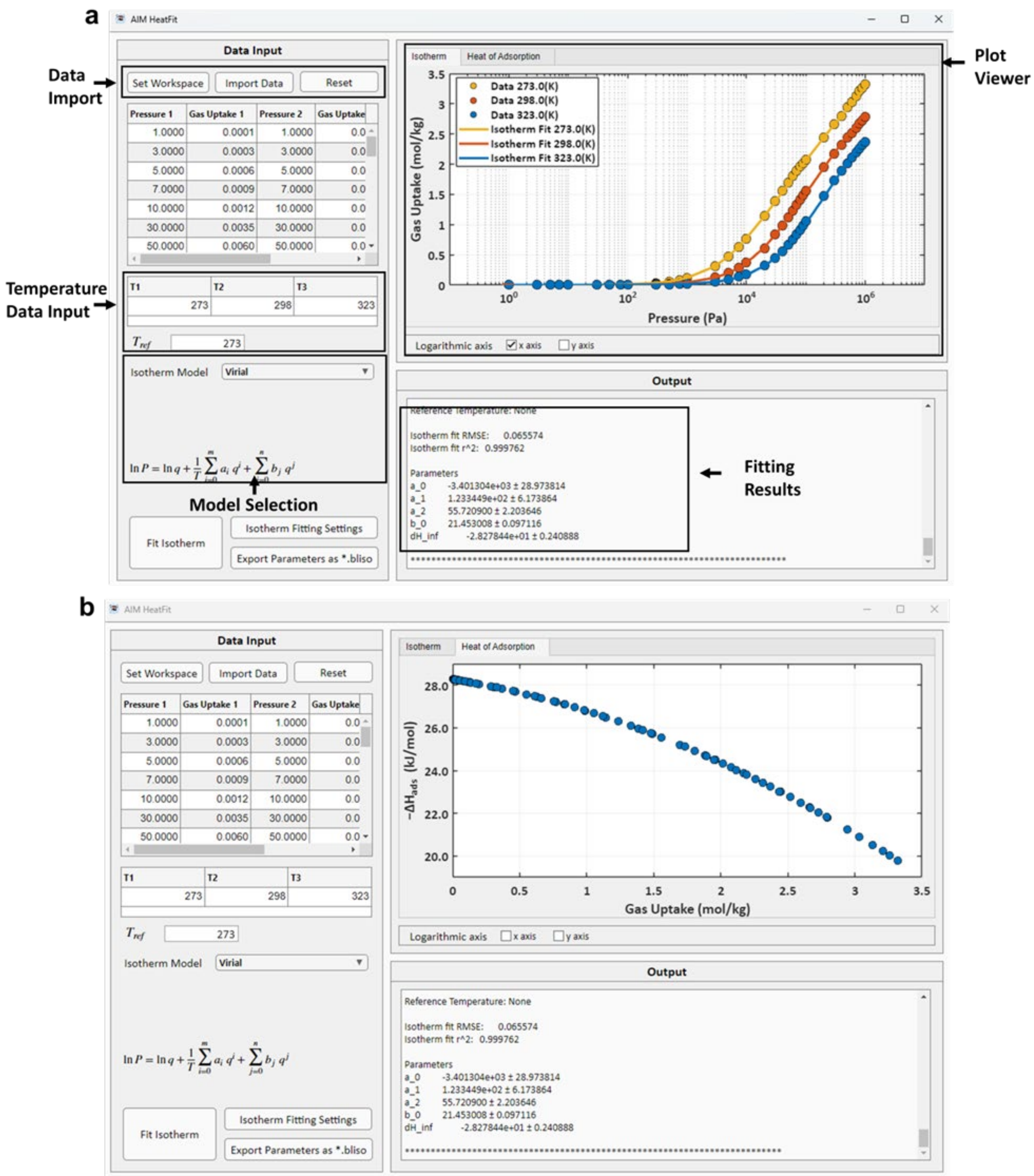

**Fig. 5.** HeatFit GUI with main isotherm fitting interface. **Data Input:** access to loading isotherm data, temperature values corresponding to isotherm data, reference temperature used for fitting, and isotherm model selection for fitting. **Plot Viewer:** displays the fitted isotherm model curve and isotherm data. **Output:** numerical output of fitted parameters and fitting statistics. **a.** Plot

Viewer showing isotherm data and the corresponding model fitting results. **b.** Plot Viewer showing predicted isosteric enthalpy oaf adsorption with respect to gas uptake

### 4.1. Clausius-Clapeyron equation

The Clausius-Clapeyron is a thermodynamic relationship which expresses the relationship between pressure and temperature for phase change processes. Since adsorption can be considered as a thermodynamic process involving a phase change, the impact of temperature on the adsorption pressure can be calculated by Clausius-Clapeyron equation. The Clausius-Clapeyron equation is:

$$\left.\frac{\partial(\ln P)}{\partial T}\right|_q = \frac{-\Delta H_{ads}}{RT^2} \tag{25}$$

where $\Delta H_{ads}$ and $R$ are the isosteric enthalpy of adsorption and general gas constant, respectively. For Clausius-Clapeyron implementation in HeatFit, we assumed that the $\Delta H_{ads}$ is constant for different temperatures and pressure, yielding the following form of Clausius-Clapeyron equation,

$$\ln\left(\frac{P_{ref}}{P}\right) = \frac{-\Delta H_{ads}}{R}\left(\frac{1}{T} - \frac{1}{T_{ref}}\right) \tag{26}$$

where $P_{ref}, T_{ref}$, and $P, T$ are the adsorption pressures and temperatures for the reference state, and given state, respectively. Both the reference state and the given state correspond to a constant adsorption loading $q$,

$$q = f(T, P) = f(T_{ref}, P_{ref}) \tag{27}$$

where $f(P, T)$ is any isotherm model. Assuming $f_{ref}$ as the isotherm model fitted at $T_{ref}$, which is only the function of $P$, we have

$$q = f(T, P) = f_{ref}(P_{ref}) \tag{28}$$

Substituting $P_{ref}$ value using equation (26),

$$q = f(T, P) = f_{ref}\left(P \times \exp\left(\frac{-\Delta H_{ads}}{R}\left(\frac{1}{T} - \frac{1}{T_{ref}}\right)\right)\right) \tag{29}$$

The utility of equation (29) is that it allows us to predict the loading at different temperatures and pressures using the isotherm model fitted at reference conditions. Additionally, if we have pressure-loading data at different temperatures, and an isotherm model fitted at the reference conditions we can also fit for $\Delta H_{ads}$ value which is averaged over the adsorption loading range.

### 4.2. Virial equation

The Virial equation is [54]:

$$\ln P = \ln q^* + \frac{1}{T}\sum_{i=0}^{m-1} a_i (q^*)^i + \sum_{j=0}^{n-1} b_i (q^*)^j \tag{30}$$

where $a_i$, and $b_i$ are the fitting parameters also known as Virial coefficients. $m$ and $n$ refers to the number of Virial coefficients $a_i$ and $b_i$, respectively. The isosteric heat of adsorption as a function of the loading is:

$$Q_{st}(q^*) = -R \sum_{i=0}^{m-1} a_i (q^*)^i \tag{31}$$

where $Q_{st}$ is the isosteric heat of adsorption, $R$ is the general gas constant. The isosteric heat of adsorption for the infinite dilution (i.e. for very low adsorption loading) is expressed as,

$$Q_{st}^0(q^*) = -R\, a_0 \tag{32}$$

where $Q_{st}^0$ is the isosteric heat of adsorption at infinite dilution, and $a_0$ is the first Virial coefficient. $\Delta H_{ads}$ and $\Delta H_{ads}^0$ can then be obtained from $Q_{st}$ and $Q_{st}^0$ using the relationship between isosteric heat of adsorption and isosteric enthalpy of adsorption given by equation (24).

## 5. MixPred – Mixture adsorption loadings

Adsorption isotherm of mixture depends on the pressure, temperature, and the mixture composition. MixPred module estimates the mixture adsorption loading for a given pressure and composition (**Figure 6**). The isotherm models fitted using IsoFit module can be directly loaded in the MixPred module, and the user can specify the desired pressure range and mixture composition to calculate the mixture loading. The calculated mixture loadings and adsorbed mole fraction are plotted in the Plot viewer and are also reported in the Output panel. The user can also export the calculated mixture loadings in several available file formats.

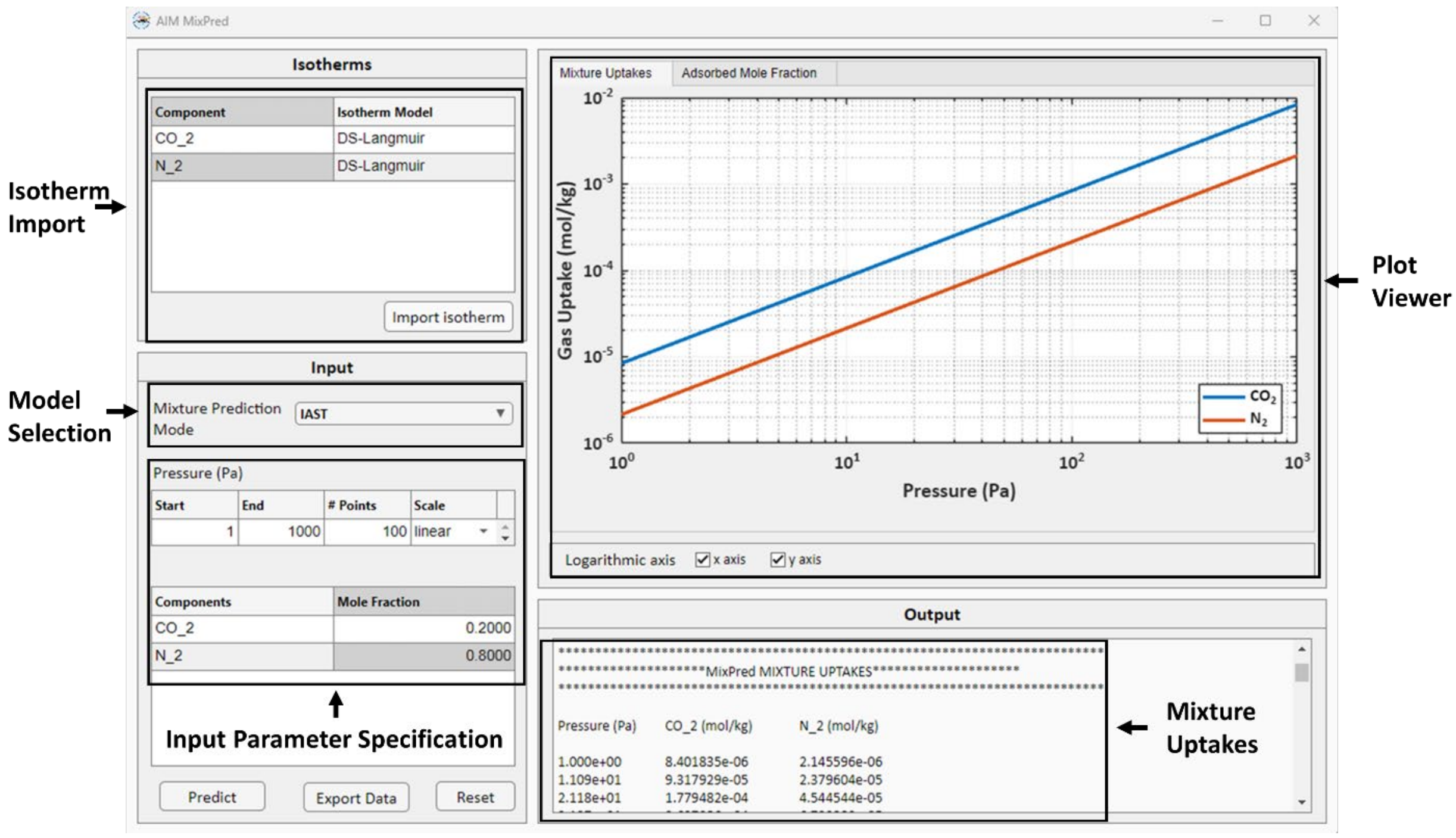

**Fig. 6.** MixPred GUI with main interface. **Isotherms**: access to load isotherm models fitted using IsoFit and HeatFit for different components. **Input**: specify mixture adsorption mode (extended model or IAST), probe pressure points for mixture adsorption loadings, and the mixture composition. **Plot Viewer:** displays the calculated mixture loadings and adsorbed mole fractions. **Output**: outputs the calculated mixture loadings

### 5.1. Extended Dual-site Langmuir model (EDSL)

The extended Langmuir model [55,56] is a generalization of the Langmuir model for multi-component adsorption. The extended Langmuir isotherm equation is:

$$q_i^* = \frac{q_{sat,1,i} b_{1,i} P_i}{1 + \sum_1^N b_{1,j} P_j} + \frac{q_{sat,2,i} b_{2,i} P_i}{1 + \sum_1^N b_{2,j} P_j} \tag{33}$$

where $N$ is the number of components involved; $q_{sat,1,i}, q_{sat,2,i}, b_{1,i}, b_{2,i}$, and $P_i$ are the Langmuir isotherm parameters and partial pressure of component $i$ respectively. The Langmuir isotherm parameters for individual components can be obtained by fitting Langmuir isotherm model to the isotherm data for the given component. The Extended Langmuir model is thermodynamically consistent only when the saturation capacities $q_{sat,1}$ and $q_{sat,2}$ for each component are equal [57], as follows:

$$q_{sat,1,1} = q_{sat,1,2} = q_{sat,1,3} = \cdots\ q_{sat,1,N} \tag{34}$$

$$q_{sat,2,1} = q_{sat,2,2} = q_{sat,2,3} = \cdots\ q_{sat,2,N} \tag{35}$$

**5.2. Ideal Adsorption Solution Theory (IAST)**

IAST is a thermodynamic framework proposed by Mayers and Prausnitz [18] to calculate the mixture isotherms from single-component isotherms. The IAST is based on three fundamental assumptions:

- The surface area of the adsorbent is equally accessible to all adsorbates.
- The adsorbed phase is an ideal solution.
- The adsorbent is homogeneous.

The solution of the IAST involves solving non-linear equations consisting of a reduced grand potential. For the component $i$, the reduced grand potential $\psi_i^*$ is given by [58]

$$\psi_i^* = \int_0^{P_i^*} \frac{q_i^*(P)}{P} dP \tag{36}$$

where $P_i^*$ is the fictitious pressure, and $q_i^*$ is the equilibrium loading of component $i$. The fictitious pressure, $P_i^*$ is the pressure for component $i$ at which it exerts the same reduced grand potential as the other components and is related to the partial pressure of component $i$, which is given as:

$$P_i = x_i P_i^* \quad \text{for } i = 1, 2, 3, \cdots, N. \tag{37}$$

where $N$ is the number of adsorbing components; $P_i$ and $x_i$ are the partial pressure and adsorbed mole fraction of component $i$, respectively. IAST states that thermodynamic equilibrium is achieved when the reduced grand potential of every component becomes equal.

$$\psi_1^* = \psi_2^* = \psi_3^* = \psi_4^* = \cdots\ \psi_N^* \tag{38}$$

In addition, the sum of the adsorbed mole fraction must be equal to 1:

$$\sum_i x_i = 1 \tag{39}$$

Together, equations (36), (37), (38), and (39) form a set of $2N$ equations that can be solved for $2N$ unknowns ($P_i^*$ and $x_i$). The equations in the IAST framework are highly non-linear and often extremely sensitive to the initial conditions [59]. Moreover, the solution is often computationally expensive and time-consuming, especially in breakthrough or cyclic simulations where IAST equations are solved many times because of the dynamic conditions inside the column. It is recommended for the user to use EDSL for faster calculation provided that the user is able to obtain sufficiently good fit of their isotherm data ($r^2$ > 0.99) over the range of operating conditions since the simulation results may be sensitive to the fitting results.

## 6. BreakLab - Breakthrough simulation

BreakLab module can be used to simulate isothermal/non-isothermal multicomponent breakthrough curves for fixed-bed systems up to 5 components (1 non-adsorbing and 4 adsorbing components). The breakthrough simulation is run for the given set of isotherm model, and parameters, and the resulting breakthrough curves are displayed (**Figure 7**). The user can interact with the simulation results using the post-processing features and can also save the resulting breakthrough data in several file formats. The module also includes functionality to save and load the simulation parameter values as a JSON file, allowing the users to efficiently store and retrieve simulation configurations. The feature reduces the need for repetitive manual input of parameter values, thereby streamlining the breakthrough simulation setup. This allows for parametric analysis which enables the users to quickly load the simulation configuration followed by the manual input of targeted parameter values and for data sharing amongst researchers to ensure consistency and reproducibility of the simulation results.

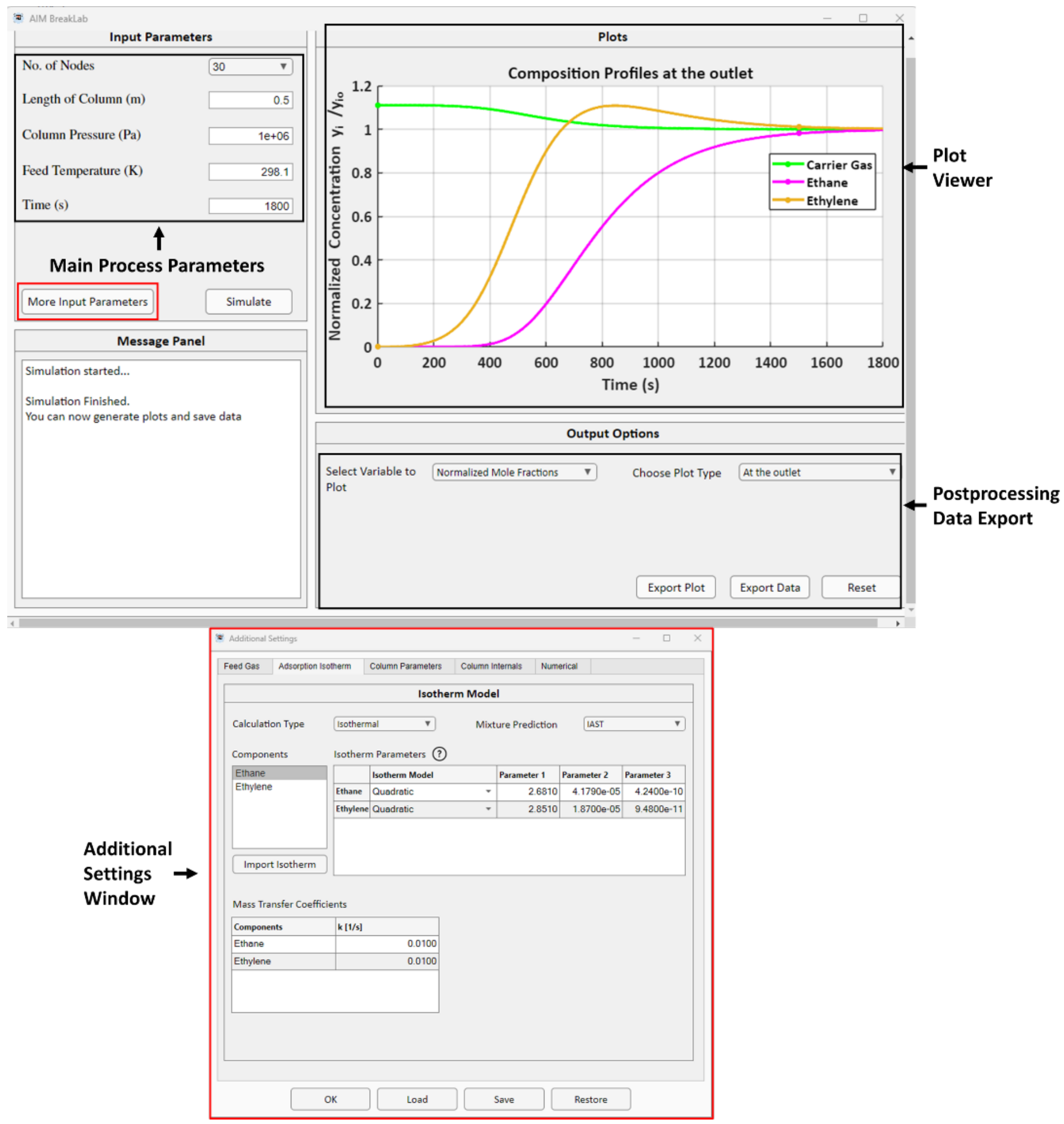

**Fig. 7.** BreakLab (GUI) showing the main breakthrough simulation interface and Additional Settings Window. **Input Parameters**: access to some key process parameters for quick simulation. **Message Panel**: reports simulation messages and any error encountered during the simulation. **Plots**: displays the breakthrough simulation results. **Output Options**: post-processing and data export options for breakthrough results. **Additional Settings Window**: access to the complete set of required simulation parameters can be opened by clicking "More Input Parameter" button. Provides options to manually input, load, and save the simulation parameters from and to JSON file

### 6.1. Model development

The mathematical model describing non-isothermal, non-adiabatic, and non-isobaric fixed bed adsorption phenomena comprises of material, momentum, and energy balances. The assumptions used in the model development are:

- The flow is axially dispersed and characterized by an axial dispersion coefficient.
- The gas phase is ideal.
- Pressure-drop across the column is given by the Ergun equation.
- The steady-state momentum balance is applicable.
- The adsorbent bed is uniform; the bed density, void fraction, and particle size are constant throughout the column.
- Thermal equilibrium exists between the solid and gas phase; gas and solid phase temperature is the same.
- The heat transfer coefficient governing the heat transfer between the bed and wall of the column is constant.
- The mass transfer resistance between the solid and gas phases is governed by the Linear Driving Force (LDF) model.
- The gradients in the radial direction are negligible.

The material, energy, and momentum balances are developed based on the conservation of mass, energy, and momentum in the fixed bed, respectively. These balance equations consist of the mass conservation of component $i$, overall mass balance, conservation of the total momentum and the total energy of the system. The balance equations are first simplified based on the assumptions listed above. The mass balance based on the concentration of component $i$ in the gas phase is given as follows:

$$\varepsilon_t \frac{\partial C_i}{\partial t} = \varepsilon_b \frac{\partial}{\partial z}\left(D_{ax} \frac{\partial C_i}{\partial z}\right) - \varepsilon_b \frac{\partial}{\partial z}(v C_i) - \rho_{b,ads} \frac{\partial q_i}{\partial t} \tag{40}$$

where $C_i$ is the concentration of component $i$; $\varepsilon_t$ , $\varepsilon_b$, $D_{ax}$, and $v$ are the total porosity, bed porosity, axial dispersion coefficient, and interstitial velocity, respectively; $\rho_{b,ads}$ is the bulk density of adsorbent, and $q_i$ is the adsorbed molar loading of component $i$. The left term in equation (40) represents the accumulation of component $i$ in the gas phase, while the first, second, and third terms on the right side represent the axial diffusion, convection, and adsorption phenomena, respectively.

The axial dispersion coefficient $D_{ax}$ is calculated using the following correlation [1]:

$$D_{ax} = 0.7D_m + v_0 r_p \tag{41}$$

where $D_m$, $v_0$, and $r_p$ are the molecular diffusivity of the feed, superficial velocity of the feed, and particle radius, respectively.

The total porosity $\varepsilon_t$ is calculated as:

$$\varepsilon_t = \varepsilon_b + (1 - \varepsilon_b)\varepsilon_p \tag{42}$$

where $\varepsilon_p$ is the particle porosity. **Figure 8** illustrates the relationship between different types of porosities.

Based on the ideal gas assumption and Dalton's law of partial pressures, the gas phase concentrations $C_i$ can be expressed in terms of partial pressure, $P_i$:

$$C_i = \frac{P_i}{RT} = \frac{y_i P}{RT} \tag{43}$$

where $y_i$ is the mole fraction of component $i$, $P$ and $T$ are the total pressure and temperature of the gas phase, respectively, and $R$ is the gas constant. The material balance for component $i$ can then be expressed in terms of mole fractions using equation (43).

$$\varepsilon_t \left( \frac{\partial y_i}{\partial t} + \frac{y_i}{P}\frac{\partial P}{\partial t} - \frac{y_i}{T}\frac{\partial T}{\partial t} \right) = D_{ax}\frac{\varepsilon_b T}{P}\frac{\partial}{\partial z}\left( \frac{P}{T}\frac{\partial y_i}{\partial z} \right) - \frac{\varepsilon_b T}{P}\frac{\partial}{\partial z}\left( \frac{y_i v P}{T} \right) - \frac{\rho_{b,ads} RT}{P}\frac{\partial q_i}{\partial t} \tag{44}$$

To close the material balance, an overall material balance equation can be obtained by summing equation (44) over all the components $i \in N$, yielding an explicit equation in terms of the total pressure of the system.

$$\frac{\partial P}{\partial t} = \frac{P}{T}\frac{\partial T}{\partial t} - \frac{\varepsilon_b T}{\varepsilon_t}\frac{\partial}{\partial z}\left(\frac{vP}{T}\right) - \frac{\rho_{b,ads}RT}{\varepsilon_t}\sum_{i\in I}\frac{\partial q_i}{\partial t} \tag{45}$$

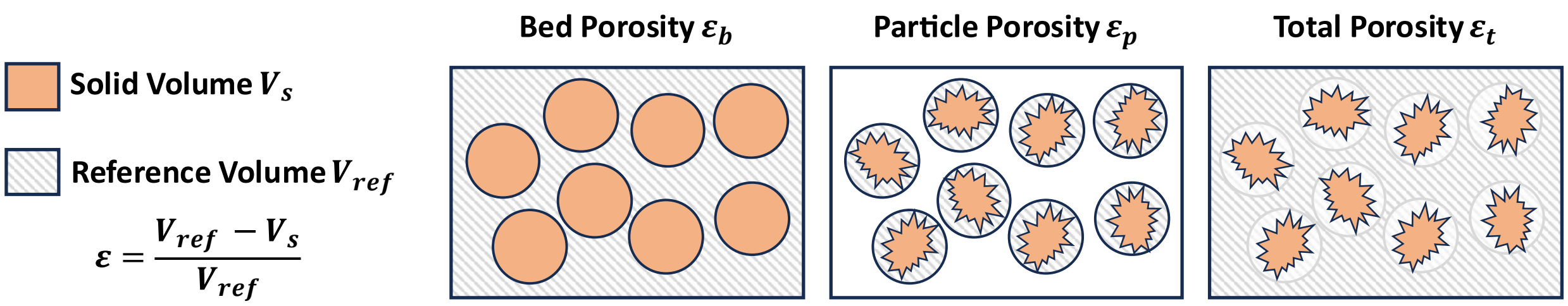

**Fig. 8.** Schematic showing the choice of reference and solid volumes to calculate different types of porosities. Bulk porosity $\varepsilon_b$ accounts for the free space outside the particles. Particle porosity $\varepsilon_p$ represents the free space within the particle. The total porosity $\varepsilon_t$ considers the free volume both outside and inside the particle

For non-isothermal breakthrough simulations, the energy balance is required to calculate temperature changes in the column to account for the heat generated from adsorption. Since the model assumes thermal equilibrium between the gas and solid phase, only gas phase energy balance is required. The energy balance for the gas phase is:

$$\begin{aligned}\left(\rho_{b,ads}C_{p,ads} + \rho_{b,ads}C_{p,a}\sum_{i\in I} q_i\right)\frac{\partial T}{\partial t} \\ = K_z\frac{\partial^2 T}{\partial z^2} - \frac{C_{p,gas}\varepsilon_b}{R}\frac{\partial}{\partial z}(vP) + \rho_{b,ads}\left(\sum_{i\in I}\left(-\Delta H_{ads,i}\frac{\partial q_i}{\partial t}\right) - C_{p,a}T\sum_{i\in I}\frac{\partial q_i}{\partial t}\right) \\ - \frac{2h_{in}}{r_{in}}(T - T_{wall}) - \frac{C_{p,gas}\varepsilon_t}{R}\frac{\partial P}{\partial t}\end{aligned} \tag{46}$$

where $K_z$ is the thermal conductivity of gas; $C_{p,ads}$, $C_{p,a}$, and $C_{p,gas}$ are the specific heat capacities of the adsorbent, adsorbed phase, and gas phase, respectively; $\Delta H_{ads,i}$ is the isosteric enthalpy of adsorption of component $i$; $h_{in}$, $r_{in}$, and $T_{wall}$ are the inside heat transfer coefficient, the inner radius of the breakthrough column, and the column wall temperature, respectively. We assume that $C_{p,a}$ is equal to $C_{p,gas}$. The terms on the right side of the equation account for the conduction, convection, heat effects due to adsorption, and heat transfer to the wall. The last term accounts for the temperature changes due to moles of gas itself.

The temperature changes in the column wall are calculated using wall energy balance given as,

$$\rho_{wall} C_{p,wall} \frac{\partial T_{wall}}{\partial t} = K_{z,wall} \frac{\partial^2 T_{wall}}{\partial z^2} - \frac{2 h_{in} r_{in}}{\left(r_{out}^2 - r_{in}^2\right)} (T_{wall} - T) - \frac{2 h_{out} r_{out}}{\left(r_{out}^2 - r_{in}^2\right)} (T_{wall} - T_a) \qquad (47)$$

where $\rho_{wall}, C_{p,wall}$, and $K_{z,wall}$ are the mass density, specific heat capacity, and thermal conductivity of column wall, respectively; $r_{out}$ and $h_{out}$ are the outer radius of the column and outside heat transfer coefficient, respectively. $T_a$ is the ambient or wall surroundings temperature. The terms on the right side of the above equation account for heat conduction in the axial direction, heat transfer to the gas and adsorbent bed, and heat transfer to the outside of the wall, respectively. In case the user of the program does not want to simulate the wall temperature changes, the wall energy balance can be decoupled from the mathematical model by specifying $K_{z,wall}, h_{in}$, and $h_{out}$ equal to zero.

Axial pressure drop across the packed bed is modelled using the Ergun equation.

$$-\frac{\partial P}{\partial z} = \left(\frac{150\mu}{4 r_p^2}\right) \left(\frac{1-\varepsilon_b}{\varepsilon_b}\right)^2 v + \left(\frac{1.75 \rho_{gas}}{2 r_p} \left(\frac{1-\varepsilon_b}{\varepsilon_b}\right)\right) v^2 \qquad (48)$$

where $\mu$, $r_p$, and $\rho_{gas}$ are the viscosity of the gas phase, particle radius, and mass density of the gas, respectively. The mass transfer between the adsorbent and gas phase is given by the linear driving force (LDF) model.

$$\frac{\partial q_i}{\partial t} = k_i (q_i^* - q_i) \qquad (49)$$

where $k_i$ is the mass transfer coefficient of adsorbing component $i$, and $q_i^*$ is the mixture equilibrium molar loading calculated by the user-defined single component isotherm model and parameters. For the mixture equilibrium molar loading, the user can choose between extended

Langmuir and IAST. The user can import single component isotherm model parameters obtained from IsoFit or HeatFit or directly choose isotherm model and specify the required parameters for the given component within the module. Users must specify the mass transfer coefficients of adsorbing species, which can be obtained experimentally or computationally.

### 6.2. Boundary conditions

**Table 3** lists the Danckwerts boundary conditions [60,61] implemented for the breakthrough simulation. Equations (42-49) along with the boundary conditions constitute the complete mathematical model to simulate breakthrough curves. **Table 4** summarizes all the physical input parameters required to run the breakthrough simulation. The module requires the user to specify the feed flow in either $\mathrm{ms^{-1}}$" or "$\mathrm{std.\ cm^3\ mol^{-1}}$". If the feed flow is specified in the units of "$\mathrm{ms^{-1}}$", the module treats the feed flow as superficial velocity of the feed $v_0$. If the feed flow is specified in "$\mathrm{std.\ cm^3\ mol^{-1}}$", the module treats the feed flow as volumetric feed flow at standard conditions ($P = 101325\ Pa\ and\ T = 273K$). The module then utilizes ideal gas law, feed gas temperature and pressure to get the volumetric flow. The volumetric flow and inner diameter of the column is then used to calculate $v_0$. Note that $T_{ref}$ and $\Delta H_{ads,i}$ are only required for non-isothermal breakthrough simulation which uses the equation (29) to calculate the gas adsorption loading for different temperatures.

**Table 3.** Boundary conditions implemented in BreakLab

| | | **Inlet boundary** | **Outlet boundary** |
|---|---|---|---|
| Mole fraction | $y_i$ | $D_{ax}\frac{\partial y_i}{\partial z}\Big\|_{z=0} = -v\|_{z=0}(y_{i,0} - y_i\|_{z=0})$ | $\frac{\partial y_i}{\partial z}\Big\|_{z=L} = 0$ |
| Temperature | $T$ | $K_Z\frac{\partial T}{\partial z}\Big\|_{z=0} = -\varepsilon_b v\|_{z=0}\rho_{gas}C_{p,gas}(T_0 - T\|_{z=0})$ | $\frac{\partial T}{\partial z}\Big\|_{z=L} = 0$ |
| Velocity | $v$ | $v\|_{z=0} = v_0$ | − |
| Pressure | $P$ | $P\|_{z=0} = f(v)$ | $P\|_{z=L} = P_0$ |
| Wall temperature | $T_{wall}$ | $T_{wall}\|_{z=0} = T_a$ | $T_{wall}\|_{z=L} = T_a$ |

**Table 4.** List of physical parameters required for breakthrough simulation in BreakLab

| Parameters | Description | Units |
|---|---|---|
| | **Feed Gas** | |
| $T_0$ | Feed gas temperature | K |
| $v_0$ | Feed Flow | $\mathrm{m\ s^{-1}}$<br>OR<br>$\mathrm{std\ cm^3\ min^{-1}}$ |
| $y_{i,0}$ | Feed gas mole fraction of component $i$ | $-$ |
| $MW_i$ | Molecular weight of component $i$ | $\mathrm{kg\ mol^{-1}}$ |
| $D_m$ | Molecular diffusivity of gas | $\mathrm{m^2\ s^{-1}}$ |
| $K_z$ | Thermal conductivity of gas | $\mathrm{W\ m^{-1}\ K^{-1}}$ |
| $C_{p,gas}$ | Specific heat capacity of gas | $\mathrm{J\ mol^{-1}\ K^{-1}}$ |
| $\mu$ | Viscosity of gas | Pa s |
| | **Fixed Bed Column** | |
| $P_0$ | Pressure | Pa |
| $L$ | Total length | m |
| $D_{col,in}$ | Inner diameter | m |
| $D_{col,out}$ | Outer diameter | m |
| $K_{z,wall}$ | Thermal conductivity of wall | $\mathrm{W\ m^{-1}K^{-1}}$ |
| $\rho_{wall}$ | Density of wall | $\mathrm{kg\ m^{-3}}$ |
| $C_{p,wall}$ | Specific heat capacity of wall | $\mathrm{J\ kg^{-1}K^{-1}}$ |
| $h_{in}$ | Inside heat transfer coefficient | $\mathrm{W\ m^{-2}\ K^{-1}}$ |
| $h_{out}$ | Outside heat transfer coefficient | $\mathrm{W\ m^{-2}\ K^{-1}}$ |
| $T_a$ | Ambient or surroundings temperature | K |
| | **Fixed Bed Column Internal** | |
| $\rho_{b,ads}$ | Adsorbent bulk density | $\mathrm{kg\ m^{-3}}$ |
| $D_p$ | Diameter of adsorbent particles | m |
| $\varepsilon_b$ | Bulk porosity | $-$ |
| $\varepsilon_p$ | Particle porosity | $-$ |
| $C_{p,ads}$ | Specific heat capacity of adsorbent bed | $\mathrm{J\ kg^{-1}\ K^{-1}}$ |
| | **Adsorption Isotherm** | |
| Isotherm Parameters | Isotherm parameters for the chosen isotherm model. | Various units |
| $\Delta H_{ads,i}$ | Isosteric enthalpy of adsorption for component $i$ | $\mathrm{kJ\ mol^{-1}}$ |
| $T_{ref}$ | Reference temperature for isotherm fitting | K |
| $k_i$ | Mass transfer coefficient of component $i$ | $\mathrm{s^{-1}}$ |

## 7. Numerical methods

In this section, we discuss the underlying numerical procedures for isotherm fitting and breakthrough simulation within the modules.

### 7.1. Isotherm fitting in IsoFit

The isotherm fitting is carried out using non-linear regression by minimizing the sum of squared error ($SSE$) function:

$$SSE = \min_{a_k} \sum_i^N \left(q_{i,exp}^* - f(P_i; \{a_k\}_1^M)\right)^2 \tag{50}$$

where $N$ and $q_{i,exp}^*$ represents the total number of data points and the gas uptake for the given data point $i$, respectively. $f(P_i; \{a_k\}_1^M)$ is the isotherm function where, $P_i$ is the pressure value for the given data point $i$, $a_k$ is the set of parameters and $M$ is the total number of parameters for the given isotherm model.

The regression is performed using the MATLAB's built-in non-linear least square solver $lsqnonlin$ [62]. The user can control the regression process by specifying custom initial guesses, as well as lower and upper bounds for the fitting parameters. Additionally, we offer a multistart option, which generates 1000 random initial guess within the specified bounds. The fitting process is then performed sequentially for each initial guess, and the best fitting result is selected and displayed. The multistart approach is useful for fitting problems with multiple parameter solutions. In such cases, fitting using multistart option can identify the global minimum corresponding to the best parameter estimates. The multistart option is available for all isotherm models except in the 'Auto' mode. Using multistart option for 'Auto' mode can be computationally expensive leading to excessive running times.

Root mean square error (RMSE) is used in the module to evaluate the goodness of fit.

$$RMSE = \sqrt{\frac{SSE}{N - M}} \tag{51}$$

where *N* and *M* are the total number of data points and the number of parameters in the isotherm model, respectively. If the user chooses the 'Auto' mode, the module reports the best isotherm model with the lowest RMSE values. If two or more models have same value of $SSE$, then the module will choose the model with a smaller number of parameters because of lower value of RMSE.

The module also reports coefficient of determination, $r^2$, value defined as:

$$r^2 = 1 - \frac{SSE}{\sum\left(q^*_{i,exp} - \overline{q^*_{i,exp}}\right)^2} \tag{52}$$

where $\overline{q^*_{i,exp}}$ is the mean value of the gas uptakes.

### 7.2. Isotherm fitting and isosteric enthalpy of adsorption prediction in HeatFit

The isotherm fitting and prediction of the isosteric enthalpy of adsorption are carried out using either the Clausius-Clapeyron or Virial equations where the fitting methods to obtain the enthalpy of adsorption values are different depending on the equations.

#### 7.2.1. Clausius-Clapeyron

The module uses the methodology devised by Ga *et al.* [26], which involves the following three steps:

- Step 1: Isotherm fitting at reference temperature
- Step 2: Fitting for Θ parameter
- Step 3: Fitting for isosteric enthalpy of adsorption $\Delta H_{ads}$

Here, the regression is performed using MATLAB built-in $lsqnonlin$ solver [62]. The user can control the regression process using different sets of values for initial guesses and also setting the lower and upper bounds of the parameters.

**Step 1: Isotherm fitting at reference temperature**

The module carries out isotherm fitting for the isotherm data at the user-specified isotherm model and reference temperature using non-linear regression. Isotherm fitting is carried out in a similar manner as described in IsoFit section (**section 7.1**). The module minimizes the *SSE* function given as,

$$SSE = \min_{a_k} \sum_i^N \left(q_{i,exp}^* - f_{ref}(P_i; \{a_k\}_1^M)\right)^2 \quad (53)$$

where $N$ and $q_{i,exp}^*$ represents the total number of data points and the experimental gas uptake for the given data point $i$ at reference temperature, respectively. $f_{ref}\left(P_{i,ref}; \{a_k\}_1^M\right)$ is the isotherm model where, $P_{i,ref}$ is the pressure value for the given data point $i$ at reference temperature, $a_k$ is the set of parameters and $M$ is the total number of parameters for the given isotherm model. RMSE and $r^2$ values, given by equations (51) and (52), respectively, are used to evaluate the goodness of fit. If the user chooses 'Auto' mode, the module selects the isotherm model corresponding to the lowest RMSE value.

**Step 2: Fitting for Θ parameter**

The isotherm model and the parameters obtained from fitting at reference temperature in step 1 are used to fit a set of parameters Θ, which is defined as,

$$\Theta = \{\theta_j\}_{j=1}^Z \quad (54)$$

where $Z$ is the total number of temperature values used in the fitting process. Every $\theta_j$ value in the Θ vector corresponds to temperature value $T_j$. Then the program uses non-linear regression to fit for each $\theta_j$ value with the objective function defined as follows,

$$SSE = \min_{\theta_j} \left(q_{j,exp}^* - g\left(P; \theta_j\right)\right)^2 \quad (55)$$

where $q^*_{j,exp}$ is the experimental gas uptake at temperature $T_j$ while $g(P;\theta_j)$ can be expressed as,

$$g(P;\theta_j) = f_{ref}\left(P \times \theta_j; \{a_{k,ref}\}_{k=1}^{M}\right) \tag{56}$$

where $f_{ref}$ and $a_{k,ref}$ represents the isotherm function and isotherm function parameters obtained in step 1 using reference conditions, respectively.

**Step 3: Fitting for $\Delta H_{ads}$**

The module uses the relationship between $\theta_j$ and $T_j$ to fit for $\Delta H_{ads}$ . The relationship is given by the Clausius-Clapeyron equation as:

$$\theta_{j,pred} = \exp\left(\frac{-\Delta H_{ads}}{R}\left(\frac{1}{T_j} - \frac{1}{T_{ref}}\right)\right) \tag{57}$$

The module uses non-linear regression to fit for $\Delta H_{ads}$ value while minimizing the objective function defined as:

$$SSE = \min_{\Delta \mathrm{H_{ads}}} \sum_{j=1}^{Z} (\theta_j - \theta_{j,pred}) \tag{58}$$

The module also reports the RMSE value for $\Delta H_{ads}$ fitting. The RMSE value is evaluated as:

$$RMSE = \sqrt{\frac{SSE}{Z}} \tag{59}$$

Once the fitting finishes, the module reports the isotherm parameters fitted at the reference temperature along with the fitted $\Delta H_{ads}$.

The built in isotherm fitting integrated with Clausius-Clapeyron method in the module enables user to fit $\Delta H_{ads}$ using different reference isotherm models. It should be noted that the Clausius-Clapeyron methodology discussed above employs a numerical approach and yields a single value of $\Delta H_{ads}$ averaged over the entire adsorption loading. This method is particularly suitable for obtaining $\Delta H_{ads}$ value for process modelling and simulation of adsorption processes such as breakthrough simulation in BreakLab where a single representative value of $\Delta H_{ads}$ is required for energy balance. In contrast, alternative tools such as IHoA [29] can determine heat of adsorption as a function of adsorption loading using both analytical and numerical approaches based on Clausius-Clapeyron method. We recognize this limitation, and it will be addressed in the future release of the software.

### 7.2.2. Virial equation

The module carries out Virial equation parameters fitting using non-linear regression. Isosteric enthalpy of adsorption as a function of adsorption loading is calculated using equations (24), (31) and (32). The fitting of Virial equation parameters is carried out in a similar manner to the isotherm model fitting procedure described in IsoFit section (**section 7.1**). The only exception is the definition of sum of squared error function which in the case of Virial equation is

$$SSE = \min_{a_k} \sum_i^N \left( \ln(P_{i,exp}) - f\left(q_{i,exp}^*, T_i; \{a_k\}_1^M\right) \right)^2 \quad (60)$$

Where $N$ is the total number of data points, $P_{i,exp}, q_{i,exp}^*$, and $T_i$ are the experimental pressure value, experimental gas uptake, and temperature values for the given data point $i$, respectively. $a_k$ and $M$ represent the $k^{th}$ Virial parameter and total number of the Virial parameters, respectively. $f$ is the Virial equation given in equation (30). This form of $SSE$ is used because the Virial equation expresses the natural logarithm of adsorption pressure as a function of adsorption uptake and temperature.

### 7.3. Numerical solution of IAST equations

AIM modules MixPred and BreakLab solve IAST equations using the modified FastIAS algorithm introduced by O'Brien and Myers [63]. The algorithm expresses the equality condition of reduced grand potential (equation 38) as a set of $N-1$ equations given as,

$$\begin{aligned}
&\psi_1(P_1^*) - \psi_N(P_N^*) = 0\\
&\psi_2(P_2^*) - \psi_N(P_N^*) = 0\\
&\cdots\\
&\psi_{N-1}(P_{N-1}^*) - \psi_N(P_N^*) = 0
\end{aligned} \tag{61}$$

Here, $N$ is the number of components. Additionally, equations (37) and (39) are combined to give the $N^{th}$ equation,

$$1 - \sum_{i}^{N} \frac{P_i}{P_i^*} = 0 \tag{62}$$

Together, equations (61) and (62) constitute a system of $N$ nonlinear equations that can be expressed in the vector form as,

$$\mathbf{G}(\boldsymbol{P}^*) = \mathbf{0} \tag{63}$$

The equations are then solved iteratively using Newton-Raphson (NR) root finding method. The $N$ variable NR method can be expressed as,

$$\boldsymbol{P}^{*(\boldsymbol{k}+\mathbf{1})} = \boldsymbol{P}^{*(\boldsymbol{k})} - \boldsymbol{\delta}^{(\boldsymbol{k})} \tag{64}$$

$$\boldsymbol{\delta}^{(\boldsymbol{k})} = \boldsymbol{\Phi}^{-1}\mathbf{G}\left(\boldsymbol{P}^{*(\boldsymbol{k})}\right) \tag{65}$$

where $\boldsymbol{\Phi}$ is the $N \times N$ Jacobian matrix given as,

$$\boldsymbol{\Phi} = \left[\boldsymbol{\phi}_{i,j}^{k}\right]_{N\times N} = \left[\left(\frac{\partial \boldsymbol{G}_i\left(\boldsymbol{P}^{*(k)}\right)}{\partial \boldsymbol{P}_j^{*(k)}}\right)\right]_{N\times N} \tag{66}$$

Due to the structure of equations (61) and (62), Φ is sparse with non-zeros elements only in the diagonal, last column, and last row. The non-zero elements of $\boldsymbol{\Phi}$ can be calculated using the definition of reduced grand potential equation (36).

$$\begin{aligned} \phi_{i,i} &= \frac{q_i^*(P_i^*)}{P_i^*} \qquad i = 1, \dots, N-1 \\ \phi_{i,N} &= \frac{q_N^*(P_N^*)}{P_N^*} \qquad i = 1, \dots, N-1 \\ \phi_{N,i} &= \frac{P_i}{(P_i^*)^2} \qquad i = 1, \dots, N \end{aligned} \tag{67}$$

The Φ matrix can be further simplified by using equations (68) and (69). The reduced form of Φ contains non-zero elements in the diagonal and last column. The simplification transforms the matrix to upper-triangular, rendering calculation of $\delta^k$ by back substitution.

$$\phi_{N,N}^{new} = \phi_{N,N}^{old} - \sum_{j=1}^{N-1} \frac{\phi_{N,j}}{\phi_{j,j}}\, \phi_{j,N} \tag{68}$$

$$G_N^{new} = G_N^{old} - \sum_{j=1}^{N-1} \frac{\phi_{N,j}}{\phi_{j,j}}\, G_j \tag{69}$$

The solution for $\delta$ simply follows from equation (65),

$$\delta_N = \frac{G_N^{new}}{\phi_{N,N}^{new}} \tag{70}$$

$$\delta_i = \frac{G_i - \delta_N \phi_{i,N}}{\phi_{i,i}} \quad i = 1, \dots, N-1 \tag{71}$$

The modified FastIAS algorithm is fast, intrinsically robust, and converges unless physically unrealistic values (negative values of $P_i^*$) are obtained [59]. Therefore, a check is required to prevent the $P_i^*$ from becoming negative during successive iterations. We have used the following check in BreakLab and MixPred [63],

$$P_i^{*(k+1)} = \frac{P_i^{*(k)}}{2} \quad \text{if } P_i^{*(k)} - \delta^k < 0 \tag{72}$$

The NR method requires initial guesses and a suitable convergence criterion to start and terminate iterations, respectively. O' Brien and Myers [63]suggested the following initial guess,

$$P_i^{*(0)} = \frac{\sum_i^N H_i P_i}{H_i} \tag{73}$$

where $H_i$ is the Henry's coefficient of component $i$.

Later Mangano *et al*. [59] identified that the initial guess from equation (73) are much higher than actual solution for weakly adsorbing components, leading to excessive iterations and slow convergence. Mangano et al suggested the following initial guesses [59],

$$P_i^{*(0)} = \min\left(\frac{\sum_i^N H_i P_i}{H_i};\ P_T\right) \tag{74}$$

where $P_T$ is the total pressure. The upper bound of the initial guess in equation (74) is thus limited to total pressure of the system. Both MixPred and BreakLab use the initial guess described by equation (74)

O' Brien and Myers [63] used the following convergence criterion in the original FastIAS algorithm implementation,

$$\sum_{i}^{N}\left|\frac{\delta_i^k}{P_i^{*(k+1)}}\right| < 10^{-10} \tag{75}$$

However, we found out that the convergence criterion given by equation (75) is difficult to achieve when the $P_i^*$ values are small. This condition often arises at the start of breakthrough simulation in BreakLab, when the partial pressures of the components are very low, leading to low reduced grand potentials $(\psi_i)$ and fictitious pressures $(P_i^*)$. Therefore, both MixPred and BreakLab use the standard deviation of $\psi_i$ as the convergence criterion with the tolerance value of $10^{-13}$. The same convergence criterion has also been used by Sharma *et al*. in RUPTURA [25].

Once the convergence is achieved, the adsorbed mole fractions $x_i$ are calculated using equation (37). The total loading is calculated using equation (76),

$$q_{tot} = \frac{1}{\sum_i^N \left(\frac{x_i}{q_i^*(P_i^*)}\right)} \tag{76}$$

Finally, mixture loadings of components are calculated using equation (77)

$$q_{i,mix}^* = x_i q_{tot} \tag{77}$$

## 7.4. Numerical solution in BreakLab

BreakLab module uses, the approach developed by Leperi *et al.* [64] and later extended by Yancy-Caballero *et al.* [65] to solve the systems of coupled model PDEs. This work extended the number of components from two to five, implemented different isotherm models, and integrated the solver for the IAST equations.

The model equations are nondimensionalized to resolve the steep gradients in the adsorption processes and for faster convergence. The non-dimensional variables used for nondimensionalization are as follows:

$$\bar{P} = \frac{P}{P_0}, \quad \bar{T} = \frac{T}{T_0}, \quad \bar{v} = \frac{v}{v_0}, \quad x_i = \frac{q_i}{q_{i0}}$$
$$\xi = \frac{z}{L}, \quad \tau = \frac{tv_0}{L}, \quad \bar{T}_{wall} = \frac{T_{wall}}{T_0}, \quad \overline{T_a} = \frac{T_a}{T_0} \tag{78}$$

where $\bar{P}, \bar{T},$ and $\bar{v}$ are the dimensionless gas phase pressure, temperature, and interstitial velocity respectively; $x_i$ is dimensionless loading of species $i$; $\xi$ is the dimensionless length and $\tau$ is the dimensionless time; $\overline{T_{wall}}$ and $\overline{T_a}$ are the dimensionless column wall and ambient temperature respectively.

The dimensionless model equations are derived using the scaled variables in equation (78). **Table 5** lists the dimensionless model equation and parameters. Spatial discretization of the model equations is performed using the finite volume method (FVM) [66] and the weighted essentially non-oscillatory (WENO) scheme [67]. Spatial discretization converts the PDEs into a system of coupled ordinary differential equations ODEs. The module uses MATLAB's built-in stiff ODE solver ode15s [68] for solving the system of ODEs.

**Table 5.** Dimensionless model equations and groups used in BreakLab mathematical model

**Component mass balance**

$$\frac{\partial y_i}{\partial \tau} = \frac{\varepsilon_b}{\varepsilon_t Pe}\frac{\bar{T}}{\bar{P}}\frac{\partial}{\partial \xi}\left(\frac{\bar{P}}{\bar{T}}\frac{\partial y_i}{\partial \xi}\right) - \frac{\varepsilon_b}{\varepsilon_t}\frac{\bar{T}}{\bar{P}}\frac{\partial}{\partial \xi}\left(\frac{y_i \bar{v}\bar{P}}{\bar{T}}\right) - \psi_{ads}\frac{\bar{T}}{\bar{P}}\frac{\partial x_i}{\partial \tau} - \frac{y_i}{\bar{P}}\frac{\partial \bar{P}}{\partial \tau} + \frac{y_i}{\bar{T}}\frac{\partial \bar{T}}{\partial \tau}$$

**Overall mass balance**

$$\frac{\partial \bar{P}}{\partial \tau} = \frac{\bar{P}}{\bar{T}}\frac{\partial \bar{T}}{\partial \tau} - \frac{\varepsilon_b \bar{T}}{\varepsilon_t}\frac{\partial}{\partial \xi}\left(\frac{\bar{v}\bar{P}}{\bar{T}}\right) - \bar{T}\psi_{ads}\sum_{i\in I}\frac{\partial x_i}{\partial \tau}$$

**Energy balance**

$$\frac{\partial \bar{T}}{\partial \tau} = \Omega_1 \frac{\partial^2 \bar{T}}{\partial \xi^2} - \Omega_2 \frac{\partial}{\partial \xi}(\bar{v}\bar{P}) + \sum_{i\in I}\left(\sigma_{i,ads}\frac{\partial x_i}{\partial \tau}\right) - \Omega_3 \bar{T}\sum_{i\in I}\frac{\partial x_i}{\partial \tau} - \Omega_4(\bar{T} - \bar{T}_{wall}) - \Omega_5 \frac{\partial P}{\partial t}$$

**Wall energy balance**

$$\frac{\partial \bar{T}_{wall}}{\partial \tau} = \pi_1 \frac{\partial^2 \bar{T}_{wall}}{\partial \xi^2} - \pi_2(\bar{T}_{wall} - \bar{T}) - \pi_3(\bar{T}_{wall} - \overline{T_a})$$

**Momentum balance**

$$-\frac{\partial \bar{P}}{\partial \xi} = \left(\frac{150\mu L v_0}{4r_p^2 P_0}\right)\left(\frac{1-\varepsilon_b}{\varepsilon_b}\right)^2 \bar{v} + \left(\frac{1.75 L v_0^2}{2 r_p T_0}\left(\frac{1-\varepsilon_b}{\varepsilon_b}\right)\left(\sum_i \frac{MW_i y_i \bar{P}}{R\bar{T}}\right)\right)\bar{v}\|\bar{v}\|$$

**Linear driving force model**

$$\frac{\partial x_i}{\partial \tau} = \alpha_i(x_i^* - x_i)$$

**Dimensionless groups**

$$Pe = \frac{L v_0}{D_{ax}}, \qquad \psi_{ads} = \frac{R T_o q_0 \rho_{b,ads}}{P_0 \varepsilon_t}, \qquad \sigma_{i,ads} = \frac{\rho_{b,ads} q_0(-\Delta H_{ads,i})}{T_0 \Psi}, \qquad \alpha_i = \frac{k_i L}{v_0}$$

$$\Omega_1 = \frac{K_z}{L v_0 \Psi}, \qquad \Omega_2 = \frac{C_{pg}\varepsilon_b P_0}{R T_0 \Psi}, \qquad \Omega_3 = \frac{C_{p,a}\rho_{b,ads} q_0}{\Psi}, \qquad \Omega_4 = \frac{2 h_{in} L}{r_{in} v_0 \Psi},$$

$$\Omega_5 = \frac{C_{pg}\varepsilon_t P_0}{R T_0 \Psi}, \qquad \Psi = \left[\rho_{b,ads} C_{p,ads} + C_{p,a}\rho_{b,ads} q_0 \sum_i x_i\right]$$

$$\pi_1 = \frac{K_{z,wall}}{\rho_{wall} C_{p,wall} L v_0}, \qquad \pi_2 = \frac{2 r_{in} h_{in} L}{\rho_{wall} C_{p,wall} v_0 (r_{out}^2 - r_{in}^2)}$$

$$\pi_3 = \frac{2 r_{out} h_{out} L}{\rho_{wall} C_{p,wall} v_0 (r_{out}^2 - r_{in}^2)}$$

## 8. Case studies using AIM modules

This section presents the case studies for isotherm fitting and breakthrough simulations performed using AIM. We begin with a case study of single-temperature isotherm fitting using IsoFit, followed by isothermal breakthrough simulations using BreakLab. Subsequently, the application of HeatFit for multi-temperature isotherm fitting and isosteric enthalpy of adsorption prediction is demonstrated. Finally, a non-isothermal breakthrough simulation is presented, along with experimental validation, to highlight the advanced simulation capabilities of BreakLab.

### 8.1. Single temperature isotherm fitting using IsoFit

To demonstrate use of IsoFit module, we considered an adsorption isotherm fitting case study for $CO_2$ in CALF-20 at 298K. The $CO_2$ adsorption isotherm data at 298K were obtained using the grand canonical Monte Carlo (GCMC) simulations. The details of the GCMC simulation can be found in **Appendix A**. The isotherm data and the fitted model using IsoFit are shown in **Figure 9a**. The isotherm exhibits an inflection point at $10^3$ Pa, indicating the presence of two distinct adsorption sites. We considered the dual-site Langmuir (DSL) isotherm model with four parameters ($q_{sat,1}$, $b_1$, $q_{sat,2}$, and $b_2$) based on the above observation. **Figure 9a** shows that IsoFit accurately fits the isotherm data. We also compared the IsoFit fitting results with other isotherm fitting codes which are RUPTURA, IAST+, and pyIAST, as shown in **Table 6**. IsoFit, RUPTURA, and pyIAST give good fit with $r^2 = 0.9990$ and RMSE $< 0.0577$. Fitting using IAST++ has a slightly higher RMSE value of $0.0626$.

For the fitted isotherm models, we consider uncertainty quantification to assess the robustness of fitting parameters given the uncertainty in the isotherm data itself. In our case, the data has been obtained from GCMC simulation. GCMC is a molecular simulation technique in which adsorption uptake is calculated by sampling the probability distribution from the grand canonical ensemble. The stochastic nature of GCMC sampling can give rise to noise in the calculated uptake. The uncertainty quantification has been performed using the Bayesian inference analysis framework described and implemented by Ward *et al.* [69]. The details about the uncertainty quantification can be found in **Appendix B**. **Figure 9b** shows the probability distribution obtained from the Bayesian inference analysis. The parameters $\mu$ and $\sigma$ are the mean and standard deviation of the parameter distributions, respectively. $\mu$ represents the best value of the parameter while $\sigma$ indicates the uncertainty associated with the model parameters. Comparing **Figure 9b** with **Table**

**6**, we observe good agreement between the isotherm parameters obtained from IsoFit and Bayesian analysis which further validates IsoFit results. In all the isotherm parameters distribution, the corresponding $\sigma$ value is order of magnitude smaller than the $\mu$ values which shows that the DSL isotherm model and the parameters obtained from IsoFit accurately describe the isotherm data.

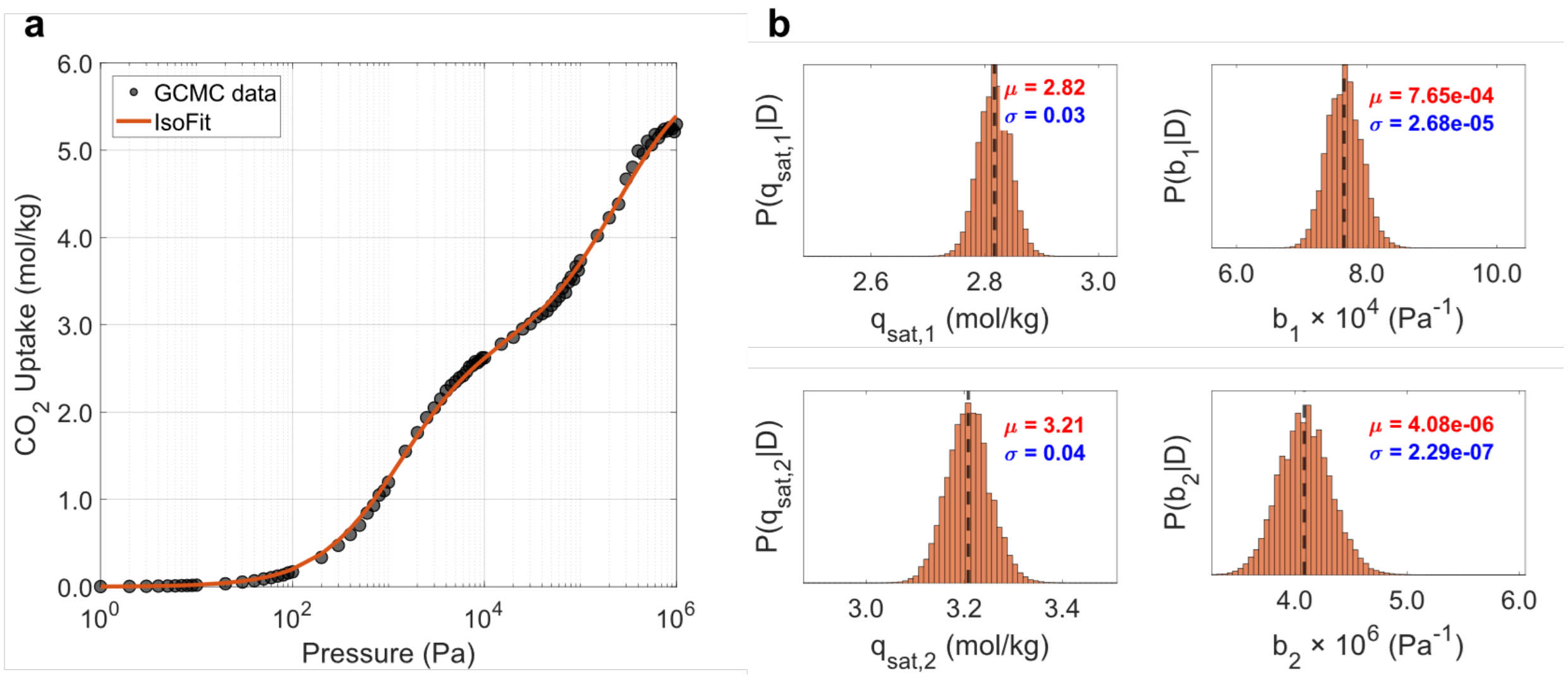

**Fig. 9.** Isotherm fitting results for $CO_2$ adsorption in CALF-20 at 298 K **a.** Dual-site Langmuir (DSL) model fit using IsoFit. Filled circles and solid lines represent the $CO_2$ uptakes from GCMC simulation and IsoFit model, respectively. **b.** Result of the uncertainty quantification for the fitted DSL model parameters. The dashed line indicates the parameter values as predicted using IsoFit. $\mu$ and $\sigma$ are the mean and standard deviation values of the corresponding parameter distributions

**Table 6.** Fitted parameters of dual-site Langmuir model (DSL) for $CO_2$ adsorption in CALF-20 at 298 K using different fitting methods.

| **Method** | $q_{sat,1}$ ($\mathrm{mol\ kg^{-1}}$) | $b_1$ ($\mathrm{Pa^{-1}}$) | $q_{sat,2}$ ($\mathrm{mol\ kg^{-1}}$) | $b_2$ ($\mathrm{Pa^{-1}}$) | $r^2$ | **RMSE** |
|---|---|---|---|---|---|---|
| IsoFit | 2.82 | $7.65\times 10^{-4}$ | 3.21 | $4.08\times 10^{-6}$ | 0.9990 | 0.0571 |
| RUPTURA | 3.21 | $4.08\times 10^{-6}$ | 2.82 | $7.65\times 10^{-4}$ | 0.9990 | 0.0571 |
| IAST ++ | 2.91 | $6.82\times 10^{-4}$ | 3.27 | $3.33\times 10^{-6}$ | 0.9988 | 0.0626 |
| pyIAST | 3.21 | $4.00\times 10^{-6}$ | 2.82 | $7.65\times 10^{-4}$ | 0.9990 | 0.0577 |

### 8.2. Isothermal breakthrough simulations using BreakLab

Two separation systems have been considered for the isothermal (no energy balance) breakthrough simulations: (1) Xe and Kr separation using SBMOF-1, and (2) $CH_4$ and $CO_2$ separation using CALF-20. In both systems, the column packed with the adsorbent is initially filled with He, which is assumed to be a non-adsorbing carrier gas. A gas mixture containing He and (1) Xe/Kr or (2) $CO_2/CH_4$ is introduced at the column inlet. The pure component adsorption isotherms at 298 K have been obtained using the GCMC simulations and fitted to Langmuir isotherm model using IsoFit module. **Figure 10** shows the GCMC data, and the isotherm fit for the two systems. **Tables 7** and **8** list the isotherm parameters for the two systems, respectively. The parameters for Xe, Kr, and $CH_4$ correspond to the single-site Langmuir (SSL) model while for $CO_2$ dual-site Langmuir (DSL) model gives the best fit. **Figure 10** shows that Xe has a higher affinity than Kr for SBMOF-1, while CALF-20 is more selective for $CO_2$ than $CH_4$.

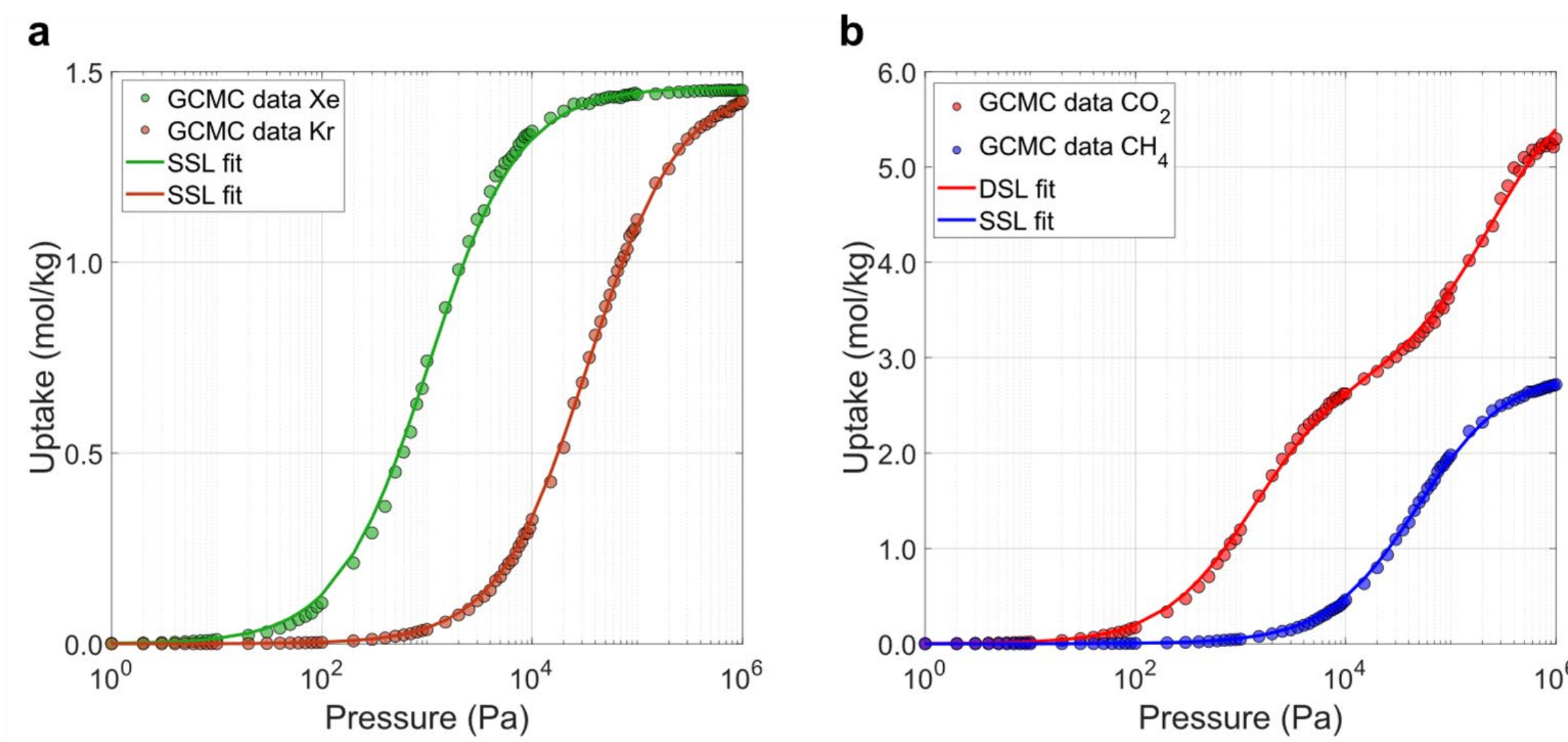

**Fig. 10.** Pure component isotherm fitting results from IsoFit: **a.** Xe/Kr adsorption in SBMOF-1, **b.** $CO_2/CH_4$ adsorption in CALF-20, with filled circles representing GCMC simulation data and solid lines showing fitting single-site (SSL) and dual-site (DSL) Langmuir models

**Table 7.** Isotherm parameters for Xe/Kr adsorption in SBMOF-1

| Parameters | Xe | Kr | Units |
|---|---|---|---|
| $q_{sat,1}$ | 1.46 | 1.47 | mol $kg^{-1}$ |
| $b_1$ | $9.68 \times 10^{-4}$ | $2.92 \times 10^{-5}$ | $Pa^{-1}$ |
| $r^2$ | 0.9992 | 0.9998 | − |

**Table 8.** Isotherm parameters for $CO_2/CH_4$ adsorption in CALF-20

| Parameters | $CO_2$ | $CH_4$ | Units |
|---|---|---|---|
| $q_{sat,1}$ | 2.82 | 2.87 | $\mathrm{mol\ kg^{-1}}$ |
| $b_1$ | $7.65 \times 10^{-4}$ | $2.08 \times 10^{-5}$ | $\mathrm{Pa^{-1}}$ |
| $q_{sat,2}$ | 3.21 | − | $\mathrm{mol\ kg^{-1}}$ |
| $b_2$ | $4.08 \times 10^{-6}$ | − | $\mathrm{Pa^{-1}}$ |
| $r^2$ | 0.9990 | 0.9994 | − |

An isothermal breakthrough simulation has been performed for both systems in BreakLab using the fitted isotherm parameters. **Table 9** lists the parameters used in the breakthrough simulation.

**Table 9.** Operating parameters for Xe/Kr and $CO_2/CH_4$ breakthrough simulations

| Parameters | | Value | Units |
|---|---|---|---|
| *Feed gas* | | | |
| Temperature | | 298.15 | K |
| Molecular diffusivity | | $1.60 \times 10^{-5}$ | $\mathrm{m^2\ s^{-1}}$ |
| Gas viscosity | | $1.72 \times 10^{-5}$ | $\mathrm{kg\ m^{-1}\ s^{-1}}$ |
| Feed velocity | | 0.04 | $\mathrm{m\ s^{-1}}$ |
| Mass transfer coefficient | | 0.06 | $\mathrm{s^{-1}}$ |
| Composition (mole fraction) | Xe: Kr: He | 0.05:0.05:0.90 | − |
| | $CO_2$: $CH_4$: He | 0.05:0.05:0.90 | − |
| *Adsorber column* | | | |
| Pressure | | $1 \times 10^6$ | Pa |
| Length | | 0.3 | m |
| Diameter | | 0.0127 | m |
| Particle diameter | | $2 \times 10^{-3}$ | m |
| Particle porosity | | 0.0 | − |
| Bed porosity | | 0.40 | − |
| Adsorbent bed density | SBMOF-1 | 942 | $\mathrm{kg\ m^{-3}}$ |
| | CALF-20 | 959 | $\mathrm{kg\ m^{-3}}$ |

**Figure 11** shows the breakthrough simulation results from both BreakLab and RUPTURA [25]. The IAST model has been used to calculate the mixture adsorption isotherm in both programs. For SBMOF-1 (**Figure 11a**), Kr breakthrough occurs before Xe because of the strong adsorption of Xe compared to Kr. Before Xe breakthrough, the Kr concentration in the outlet gas was higher than the Kr concentration in the inlet gas. This phenomenon is called roll-up, and the roll-up of the Kr composition occurs because of the lower concentration of strongly adsorbing component Xe in the gas phase and the displacement of already adsorbed Kr back into the gas phase by Xe. Therefore, the concentration of weakly adsorbing components increases in the gas phase and eventually becomes greater than the inlet concentration. Once Xe breakthrough occurs, the concentration of both Xe and Kr becomes equal to the inlet concentration. For CALF-20 (**Figure 11b**), $CH_4$ breakthrough occurred before $CO_2$, showing that CALF-20 has a greater uptake for $CO_2$ than $CH_4$. A similar roll-up behavior is also observed here for $CH_4$. **Figure 11** also shows that the results from BreakLab and RUPTURA are in excellent agreement.

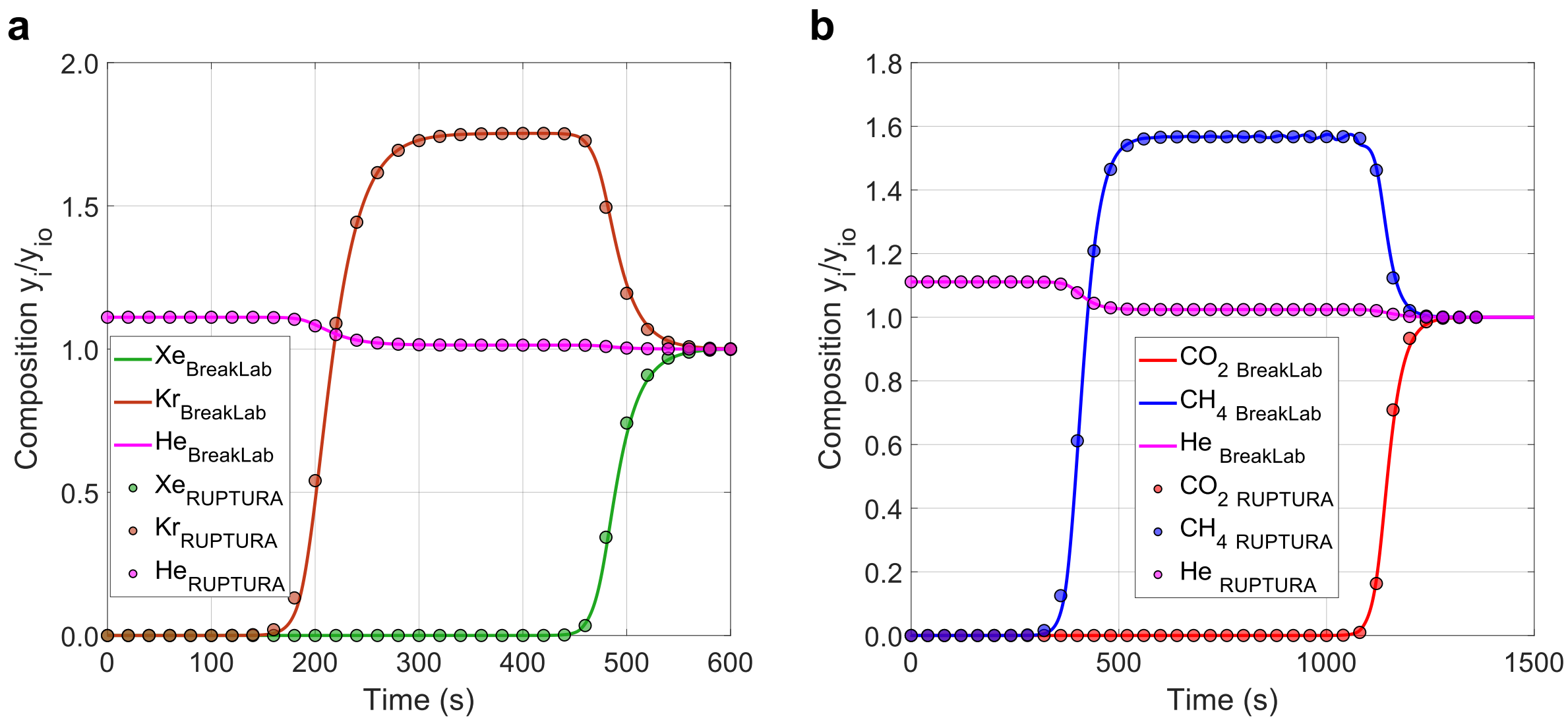

**Fig. 11.** Breakthrough curves at the outlet of the adsorber column for: **a.** Xe/Kr in SBMOF-1 and **b.** $CO_2/CH_4$ in CALF-20, comparing BreakLab (solid lines) and RUPTURA (filled circles) using IAST

The breakthrough simulation results from BreakLab using IAST and EDSL models are also compared to demonstrate the ability of BreakLab to simulate breakthrough curves using either IAST or EDSL models. **Figure 12** presents the results of such a comparison. The results from the IAST and EDSL models are nearly identical for both adsorption systems.

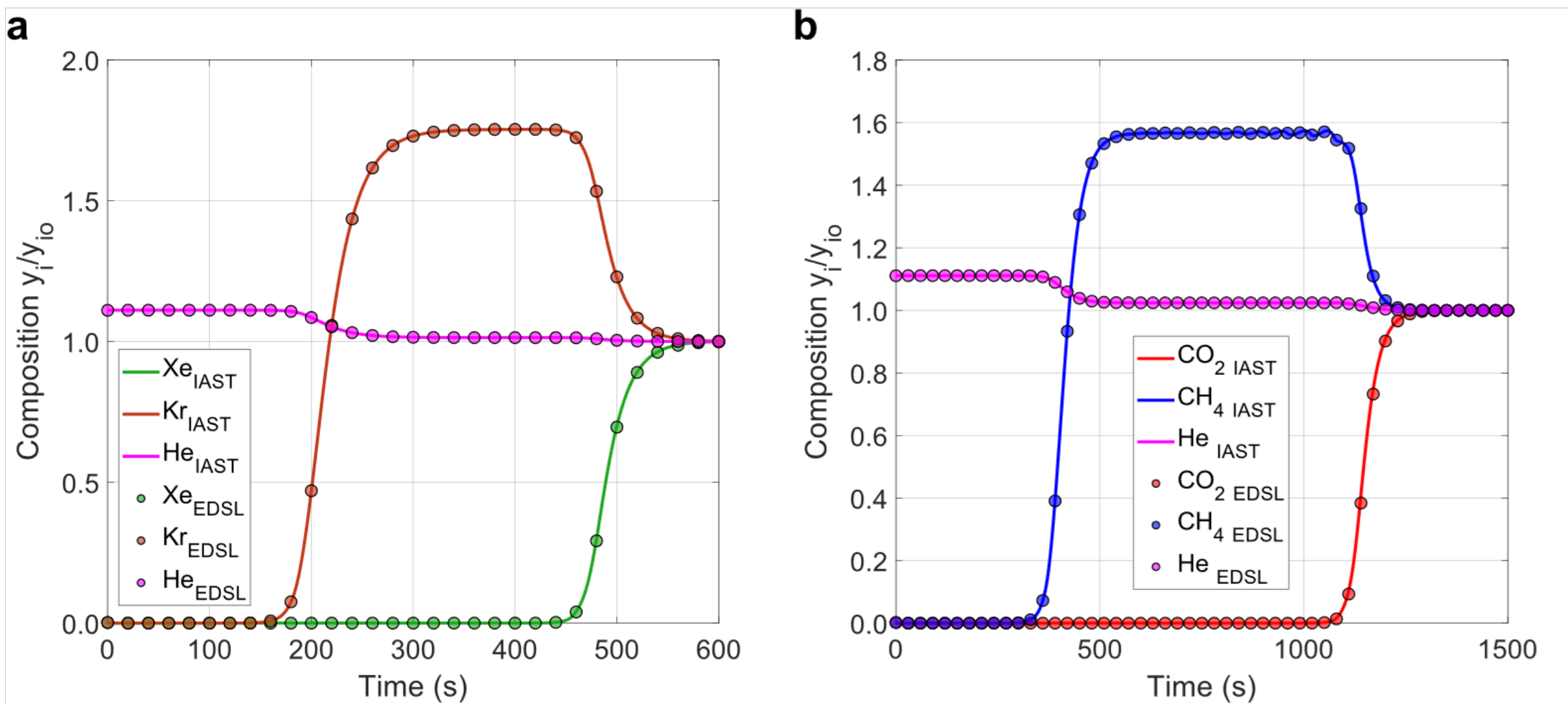

**Fig. 12.** Breakthrough curves obtained from BreakLab at the outlet of the adsorber column using the IAST and EDSL for: **a.** Xe/Kr adsorption in SBMOF-1 and **b.** $CO_2/CH_4$ adsorption in CALF-20. Filled circles and solid lines represent the breakthrough results using EDSL and IAST models, respectively

### 8.3. Estimating the Isosteric enthalpy of adsorption using HeatFit

$CO_2$ adsorption isotherms for CALF-20 were obtained at 273 K, 298 K, and 323 K using GCMC simulation and fitted using Virial equation. Fitting multi-temperature isotherm using Virial equation is often challenging because the same sets of Virial parameters $a_i$ and $b_j$ describing isotherm data at all temperatures are sought. Hence, the selection of the number of Virial parameters $a_i$ and $b_j$ is critical for ensuring fitting accuracy and reliability. Using a few numbers of parameters can lead to poor fitting while excessive number of parameters can lead to overfitting with unreliable fitting parameters. The common practice is to gradually increase the number of Virial parameters until the RMSE value does not decrease significantly.

To demonstrate the effect of the number of Virial parameters on fitting results, we fitted the Virial equation using HeatFit with: (1) 2 $a_i$ and 1 $b_j$ parameters, (2) 3 $a_i$ and 2 $b_j$ parameters, (3) 5 $a_i$ and 2 $b_j$ parameters, and (4) 7 $a_i$ and 2 $b_j$ parameters. **Figure 13** shows the isotherm data, fitting results and the predicted isosteric enthalpy of adsorption $\Delta H_{ads}$ for each case. **Table 10** summarizes the values of Virial parameters, fitting statistics, and the enthalpy of adsorption at infinite dilution $\Delta H_{ads}^{0}$(i.e. for very small $CO_2$ uptakes) calculated using equation (32) for the corresponding cases. We observe that case 1 with 2 $a_i$ and 1 $b_j$ parameters, fails to accurately fit the isotherm data, at both low and high $CO_2$ uptakes (**Figure 13a** left and middle). This leads to significant differences in HeatFit and GCMC predicted $\Delta H_{ads}$values (**Figure 13a** right).

In case 2, a better fit at low $CO_2$ uptakes was obtained using 3 $a_i$ and 2 $b_j$ parameters (**Figure 13b** middle) leading to a better agreement between GCMC and HeatFit predicted $\Delta H_{ads}$ values. A close look at equation (30) suggests that the parameters $a_0$ and $b_0$ contribute to the Virial equation most for low uptake values, while the contribution from remaining parameters becomes significant at higher uptake values. Consequently, a good fit for low uptake values gives a much better estimation of $a_0$ and $b_0$ leading to a more accurate prediction of $\Delta H_{ads}$ at low uptakes.

In case 3 (5 $a_i$ and 2 $b_j$ parameters) and case 4 (7 $a_i$ and 2 $b_j$ parameters) the fitting improved at higher $CO_2$ uptakes as well leading to an even better fit with and RMSE value of 0.2443, and 0.106, respectively. This suggests that an improved fit with lower RMSE values can always be obtained by using more Virial parameters. However, it is also important to check for the standard errors associated with the estimated parameter. **Table 10** shows that the standard errors associated with the estimated parameter values increase with the increasing number of parameters. Hence, while case 4 has the lowest RMSE value, the fitting parameters are unreliable because the standard errors for some parameters are of the same order as the corresponding parameter values (i.e $a_3$ and $a_4$ in case 4). The GUI features of HeatFit enable users to quickly try different numbers of Virial parameters for Virial equation fitting and check the RMSE value and the standard errors associated with the fitting results. We recommend that users of our program begin with a minimal number of parameters, gradually increasing the number of parameters while also keeping the check on standard errors and RMSE values until no significant improvement in RMSE is observed and the standard errors are reasonable. In the case of Virial equation fitting with many parameters, we recommend using a ‘multi-start’ feature of HeatFit to find the best possible fitting results.

**Table 10.** Virial fitting results using various number of Virial parameters for $CO_2$ adsorption in CALF-20

| Parameters | Case 1 $a_i = 2, b_j = 1$ | Case 2 $a_i = 3, b_j = 2$ | Case 3 $a_i = 5, b_j = 2$ | Case 4 $a_i = 7, b_j = 2$ | Units |
|---|---|---|---|---|---|
| $a_0$ | $-5260.6 \pm 186.4$ | $-4584.6 \pm 176.7$ | $-4583.4 \pm 144.3$ | $-4504.4 \pm 51.2$ | K |
| $a_1$ | $327.6 \pm 6.7$ | $-239.7 \pm 71.6$ | $-600.3 \pm 79.7$ | $-170.0 \pm 48.4$ | K $mol^{-1}$ |
| $a_2$ | $-$ | $41.6 \pm 2.75$ | $347.8 \pm 44.3$ | $-226.0 \pm 86.5$ | K $mol^{-2}$ |
| $a_3$ | $-$ | $-$ | $-75.2 \pm 12.7$ | $72.8 \pm 64.4$ | K $mol^{-3}$ |
| $a_4$ | $-$ | $-$ | $5.87 \pm 1.2$ | $38.6 \pm 22.3$ | K $mol^{-4}$ |
| $a_5$ | $-$ | $-$ | $-$ | $-16.4 \pm 3.6$ | K $mol^{-4}$ |
| $a_6$ | $-$ | $-$ | $-$ | $1.55 \pm 0.22$ | K $mol^{-5}$ |
| $b_0$ | $23.5 \pm 0.61$ | $21.6 \pm 0.58$ | $21.8 \pm 0.48$ | $21.4 \pm 0.17$ | $-$ |
| $b_1$ | $-$ | $1.24 \pm 0.22$ | $1.10 \pm 0.19$ | $1.27 \pm 0.067$ | $-$ |
| $\Delta H^0_{ads}$ | $-43.74 \pm 1.55$ | $-38.12 \pm 1.46$ | $-38.10 \pm 1.20$ | $-37.45 \pm 0.42$ | kJ $mol^{-1}$ |
| $r^2$ | 0.9884 | 0.9957 | 0.9972 | 0.9995 | $-$ |
| RMSE | 0.4846 | 0.2978 | 0.2421 | 0.1060 | $-$ |

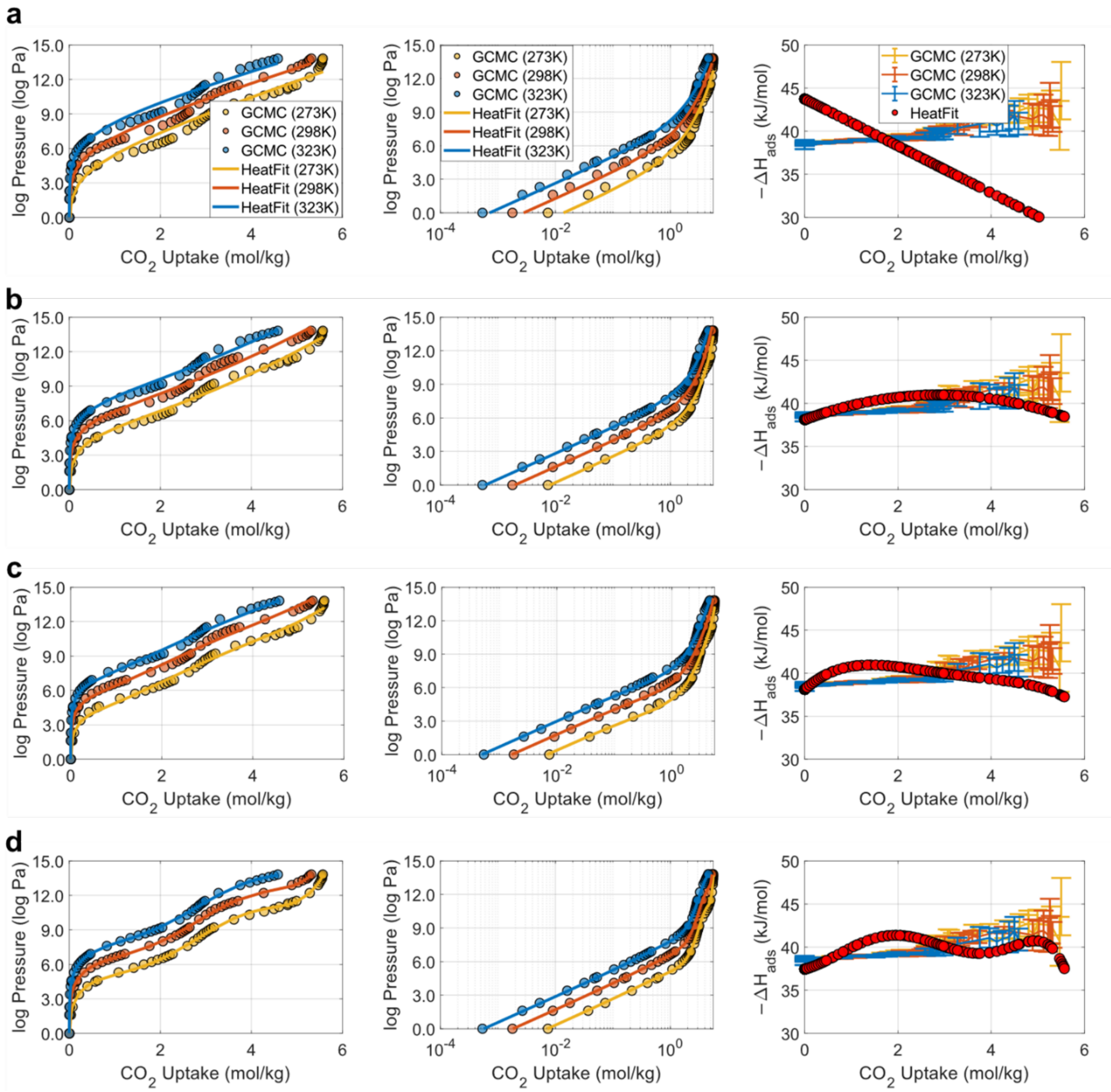

**Fig. 13.** Virial equation fitting results for $CO_2$ adsorption in CALF-20 using various number of Virial parameters. **a.** Fitting results with 2 $a_i$ and 1 $b_j$ parameters. **b.** Fitting results with 3 $a_i$ and 2 $b_j$ parameters. **c.** Fitting results with 5 $a_i$ and 2 $b_j$ parameters. **d.** Fitting results with 7 $a_i$ and 2 $b_j$ parameters. **Left.** Isotherm data and Virial fitting curves in log $p$ vs $CO_2$ uptake form. Filled circles and solid lines represent the $CO_2$ uptakes from GCMC simulation and Virial equation using HeatFit, respectively. **Middle.** Isotherm data and Virial fitting curves in log $p$ vs $CO_2$ uptake form with uptake scale in logarithmic form showing the fitting results at low $CO_2$ uptake values. **Right.** Isosteric enthalpy of adsorption $\Delta H_{ads}$ as a function of $CO_2$ uptake as obtained from GCMC

simulation and Virial equation using HeatFit. The solid lines with error bars and red filled circles correspond to $\Delta H_{ads}$ values from GCMC and Virial equation, respectively

We also consider the uncertainty quantification of Virial equation parameters for the case with 5 $a_i$ and 2 $b_j$ parameters. **Figure 14** shows the probability distributions for Virial equation parameters and isosteric enthalpy of adsorption at infinite dilution $\Delta H^0_{ads}$. The probability distributions for Virial equation parameters have been obtained from the Bayesian inference analysis discussed in **section 8.1** and **Appendix B**. The probability distribution of Virial parameter $a_0$ have been propagated through equation (32) to obtain the distribution of $\Delta H^0_{ads}$. The multiple peaks in the parameter distributions reveal that multiple combinations of these Virial parameters can describe the isotherm data. In this case, it would be misleading to calculate and rely on point estimates like mean, however, the comparison can be made by comparing the peaks of probability distribution with the parameter values obtained from HeatFit since the height of the peaks in the probability distribution represents the likelihood of the parameter values given the isotherm data. The good agreement between the parameter values obtained from HeatFit and probability distribution peaks reveals that given the uncertainty in parameter values, the fitted parameter values from HeatFit correspond to higher likelihood values. The presence of multiple peaks with varying heights as well as breadth of probability distributions implies some degree of uncertainty in estimated parameters. We also observe that the uncertainty relative to parameter values is more pronounced for higher order Virial parameters (**Table 10**). For example, the parameters $a_0$ and $b_0$ contribute to the model most for low uptake values, while the remaining parameters contribute significantly to higher uptake values and inversely proportional to the temperature. Therefore, we recommend incorporating isotherm data with higher uptake values or isotherm data measured at lower temperatures to help capture the relative influence of higher order Virial parameters on model predictions and to get a better estimate of Virial parameters.

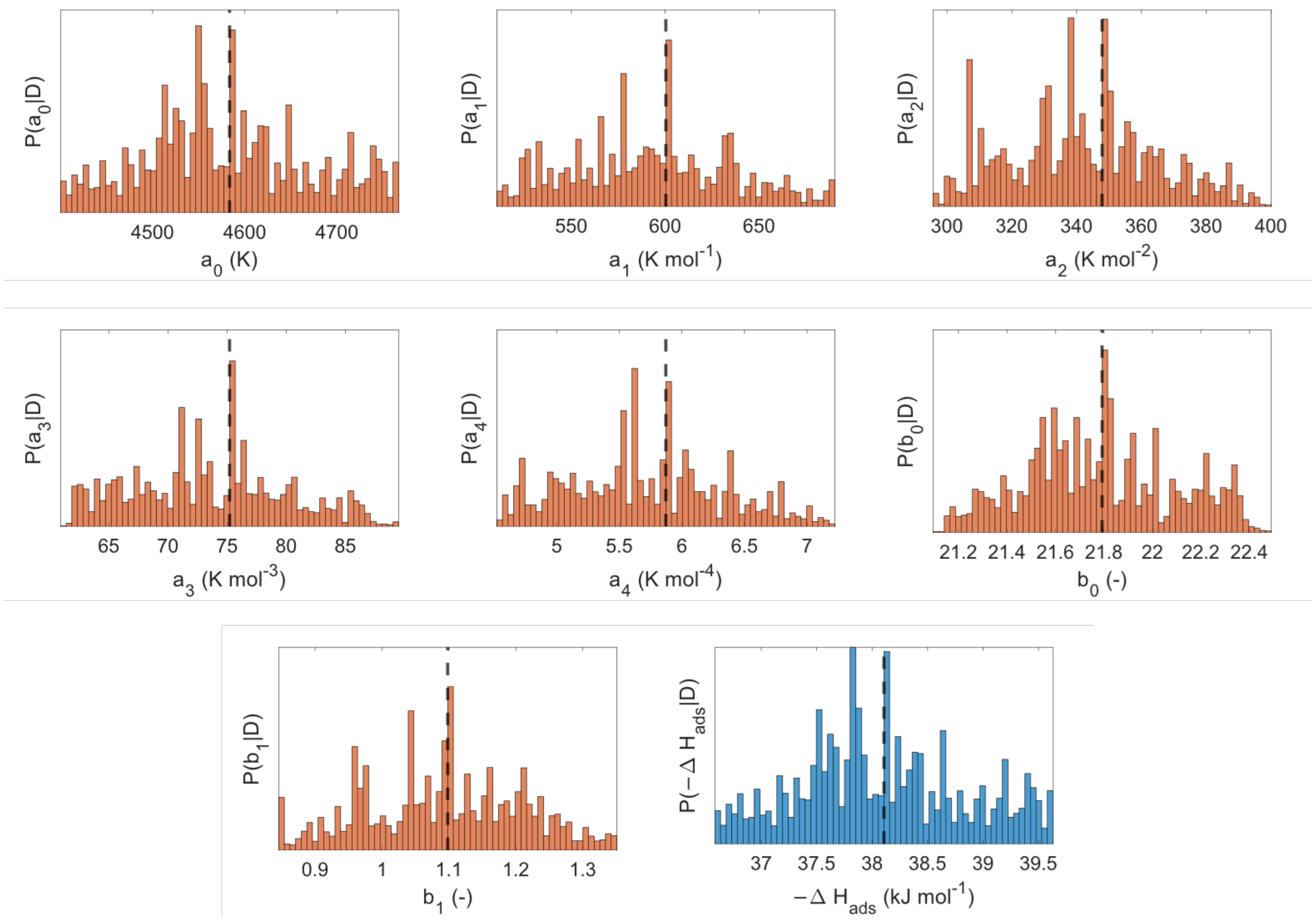

**Fig. 14.** Uncertainty quantification for the Virial equation parameters and infinite dilution isosteric enthalpy of adsorption for $CO_2$ adsorption in CALF-20. The black dashed lines indicate the fitting values obtained from HeatFit.

## 8.4. Non-isothermal breakthrough simulation using BreakLab

To validate our model, the breakthrough results from BreakLab have also been compared with the experimental breakthrough data obtained from Marx *et al*. [70]. Marx *et al.* conducted 24 experiments under varying operating conditions with ternary mixtures ($CO_2/N_2/CH_4$), and we selected a representative subset of 8 experiments for model validation. **Table 11** lists the breakthrough simulation parameters corresponding to the experimental conditions. All parameters listed in **Table 11** have been obtained directly from Marx et al. [70]. For non-isothermal breakthrough simulation, BreakLab also requires the inside heat transfer coefficient $h_{in}$ which has been estimated using the following correlation [71],

$$\frac{h_{in}D_p}{K_z} = 0.17Re^{0.79} \quad 20 \le Re \le 7600, \quad 0.05 \le \frac{D_p}{D_{col,in}} \le 0.3 \tag{79}$$

$$Re = \frac{\rho_{0,gas}v_0D_p}{\mu} \tag{80}$$

where $\rho_{gas}, v_0,$ $\mu$ and $K_Z$ are feed gas density, superficial velocity, viscosity and thermal conductivity, respectively; $D_p$ and $D_{col,in}$ are adsorbent particle diameter and fixed bed column inside diameter, respectively. The BreakLab simulation configuration JSON files for all the simulations are also available on AIM GitHub repository. IAST has been used to calculate the mixture adsorption loading.

The isotherm data for $CO_2$, $N_2$, and $H_2$ adsorption on activated carbon at 298 K, 318 K, and 338 K has been obtained from Schell *et al.* [72]. HeatFit module has been used to fit the isotherm data to Sips model and to predict the enthalpy of adsorption $\Delta H_{ads}$value. **Table 12** lists the Sips model parameters and $\Delta H_{ads}$values as obtained from HeatFit along with the reference temperature used for fitting. The reference temperature can be specified as either 298 K, 318 K, or 338 K. The reference temperatures listed in **Table 12** were selected as they provided the best fit. The isotherm parameters listed in **Table 12** correspond to the reference temperature. BreakLab uses these isotherm parameters and the reference temperature to calculate adsorption loading at different temperatures using equation (29) during the breakthrough simulation. Note that since HeatFit predicts $\Delta H_{ads}$ via Clausius-Clapeyron method when using Sips model, the $\Delta H_{ads}$ value here corresponds to an averaged value over the adsorption loading as discussed in **section 7.2.1**. **Figure 15** shows the experimental isotherm data from Schell *et al*. and isotherm fit for $CO_2$, $N_2$, and $H_2$. The RMSE values (**Table 12**) and **Figure 15** suggest good agreement between the experimental isotherm data and isotherm fits except for $CO_2$ adsorption isotherm at 338 K, where the Sips model overestimates the gas uptakes at higher pressures. The experimental data at high pressures suggests that $CO_2$ saturation capacity of activated carbon decreases with temperature. However, in HeatFit, saturation capacity is assumed independent of temperature and is obtained by fitting the isotherm data at reference temperature which may lead to inconsistency. The inconsistency in $CO_2$ uptakes can impact mixture uptakes calculated using IAST. For instance, the maximum partial pressure of $CO_2$ in the system is 4 bar (**Table 11**), which is well below the 100 bar region where the discrepancy occurs. Next, we need to consider the calculation of reduced grand potential. At higher $CO_2$ fictitious pressures, the overestimation of $CO_2$ uptake leads to a corresponding overestimation of reduced grand potential (equation 36). Consequently, $CO_2$ mixture uptake is also likely to be overestimated under these conditions. The important question is then whether the system involves higher $CO_2$ fictitious pressures or not. Note that $CO_2$ is preferentially adsorbing component in the system. IAST states that at equilibrium, reduced grand potentials of all the components are equal. Since strongly adsorbing components have a higher value of reduced grand potential than weakly adsorbing components for the same fictitious pressure, the equilibrium fictitious pressures of the weakly adsorbing components ($N_2$, $H_2$) are always greater than the equilibrium fictitious pressure of strongly adsorbing component ($CO_2$). Consequently, the reduced grand potential for $CO_2$ is less likely to be calculated at higher $CO_2$ fictitious pressures. Conversely, this also suggests that the fitting quality at higher pressures is more important for weakly adsorbing components in the system.

**Table 11.** Properties and parameters used in the non-isothermal adsorption breakthrough simulation case study [70]

| Parameters | | Value | Units |
|---|---|---|---|
| *Feed gas* | | | |
| Temperature | | 298, 318 | K |
| Specific heat capacity | | 29.51 | $J\ mol^{-1}\ K^{-1}$ |
| Thermal conductivity | | 0.028 | $W\ m^{-1}\ K^{-1}$ |
| Molecular diffusivity | | $4.3 \times 10^{-6}$ | $m^2\ s^{-1}$ |
| Gas viscosity | | $1.92 \times 10^{-5}$ | $kg\ m^{-1}\ s^{-1}$ |
| Feed velocity | | 0.020, 0.061 | $m\ s^{-1}$ |
| Feed flow | | 10, 30 | $cm^3\ s^{-1}$ |
| Composition (mole ratios) | $CO_2/N_2/H_2$ | 40/30/30 | − |
| | | 15/80/5 | − |
| Mass transfer coefficient | $CO_2$ | 0.11 (40/30/30, 298K)<br>0.15 (40/30/30, 318K)<br>0.04 (15/80/5, 298K)<br>0.07 (15/80/5, 318K) | $s^{-1}$ |
| | $N_2$ | 0.30 | $s^{-1}$ |
| | $H_2$ | 1.00 | $s^{-1}$ |
| *Adsorber column* | | | |
| Pressure | | $1 \times 10^5$, $10 \times 10^5$ | Pa |
| Length | | 1.2 | m |
| Inner diameter | | 0.025 | m |
| Outer diameter | | 0.040 | m |
| Wall thermal conductivity | | 1.0 | $W\ m^{-1}\ K^{-1}$ |
| Wall density | | $4 \times 10^6$ | $kg\ m^{-3}$ |
| Wall specific heat capacity | | 1.0 | $J\ kg^{-1}\ K^{-1}$ |
| Outside heat transfer coefficient | | 5.0 | $W\ m^{-2}\ K^{-1}$ |
| Ambient temperature | | 298 | K |
| Adsorbent bed bulk density | | 502 | $kg\ m^{-3}$ |
| Particle diameter | | $3 \times 10^{-3}$ | M |
| Particle porosity | | 0.568 | − |
| Bed porosity | | 0.403 | − |
| Adsorbent bed specific heat capacity | | 1000 | $J\ kg^{-1}\ K^{-1}$ |

**Table 12.** Isotherm parameters and enthalpy of adsorption for $CO_2$, $N_2$, and $H_2$ adsorption in activated carbon

| **Parameters** | **$CO_2$** | **$N_2$** | **$H_2$** | **Units** |
|---|---|---|---|---|
| $q_{sat}$ | 12.45 | 5.65 | 7.54 | mol $kg^{-1}$ |
| $b$ | $4.39 \times 10^{-7}$ | $2.77 \times 10^{-7}$ | $2.94 \times 10^{-8}$ | $Pa^{-1}$ |
| $n$ | 1.36 | 1.18 | 1.08 | − |
| $\Delta H_{ads}$ | −24.86 | −16.36 | −9.73 | kJ $mol^{-1}$ |
| $T_{ref}$ | 318 | 298 | 298 | K |
| $r^2$ | 0.9934 | 0.9996 | 0.9997 | − |

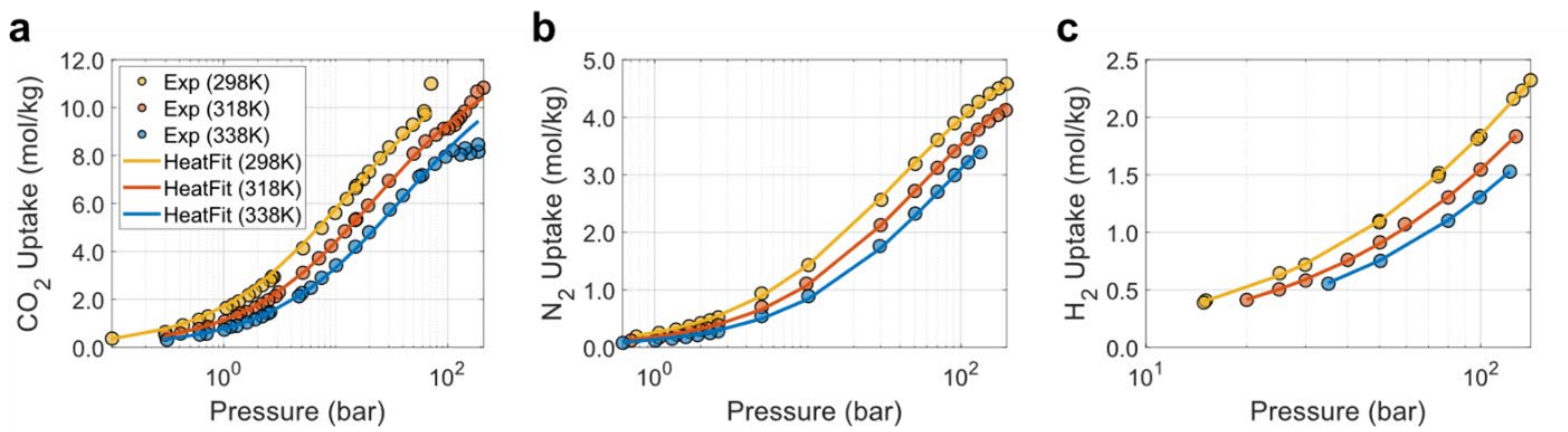

**Fig. 15.** Pure component isotherm fitting results from HeatFit at 298 K, 318 K, and 338 K for **a.** $CO_2$, **b.** $N_2$, and **c.** $H_2$, respectively. Filled circles and solid lines represent the experimental data from Schell *et al.* [72] and Sips model prediction using HeatFit, respectively.

Figure 16 and 17 shows the comparison between the breakthrough results from BreakLab and the experiment by Marx *et al*. [70] for the feed flowrate of 10 $cm^3/s$ and 30$cm^3/s$, respectively, under varying feed composition, temperature, and pressure. In the case of composition breakthrough profiles, a good agreement is observed between the simulation and experiment results for all the cases, except for the case in **Figure 16(c)** where BreakLab predicts slightly earlier breakthrough of $CO_2$ compared to the experimental data.

For temperature breakthrough profile, we note that while simulation and experiment results follow similar trends, slight discrepancies are observed for some cases when the breakthrough front of $CO_2$ exits the column. The maximum deviation between temperature values from BreakLab and experiments is around 10 ºC. This might be attributed to temperature dependence of gas properties and transport parameters which are assumed to be constant during the simulation. Other potential causes might be an in-homogeneous temperature profile along the length of the column at the start of the breakthrough experiment and significant radial temperature variations in the column. An interesting finding is that a good agreement of simulated and experimental composition breakthrough curves does not always imply a corresponding agreement in simulated and experimental temperature breakthrough curves.

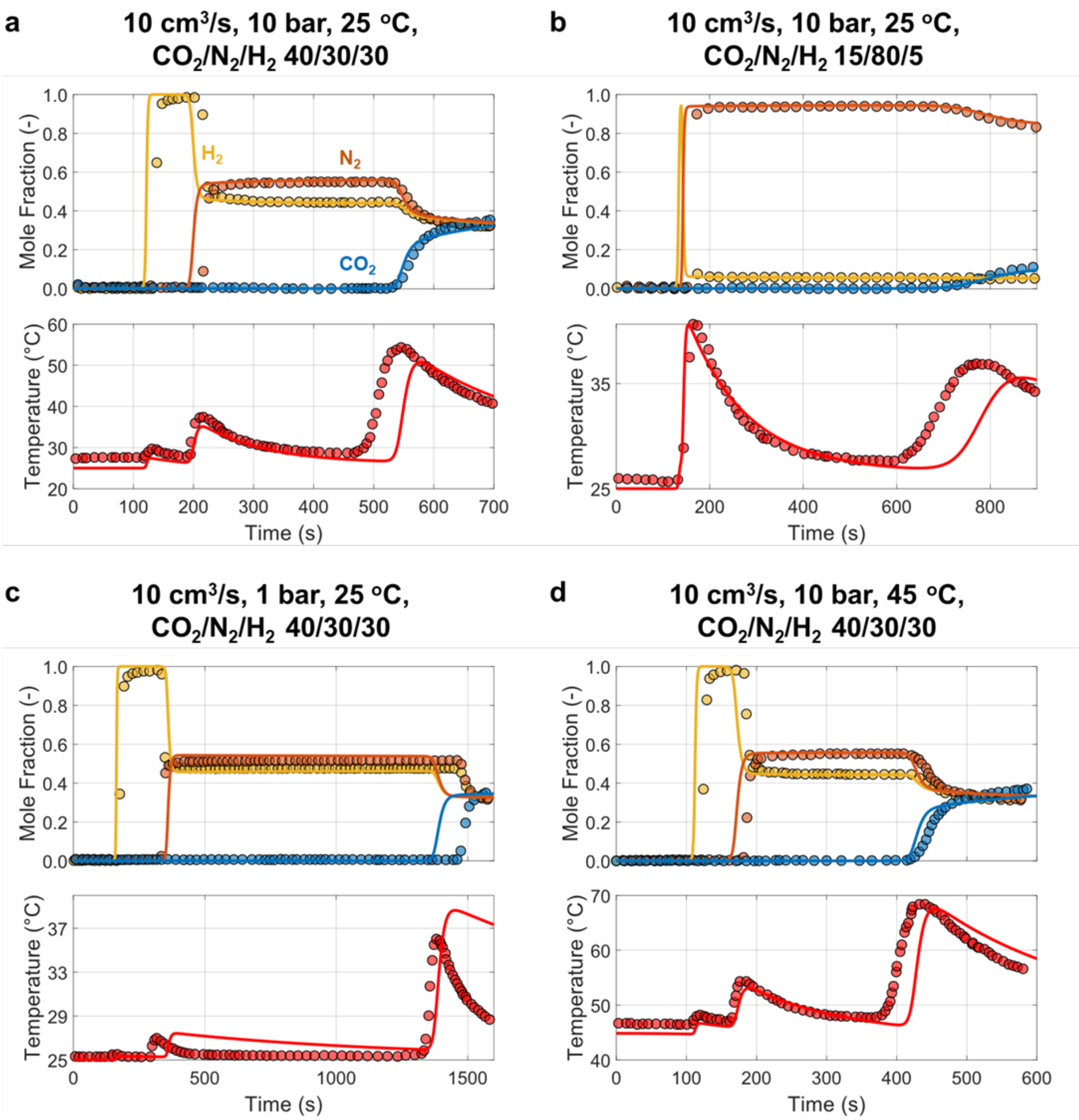

**Fig. 16.** Composition and Temperature breakthrough curves from experiments and BreakLab. Filled circles and solid lines represent the experimental data from Marx *et al.* [70] and BreakLab results, respectively

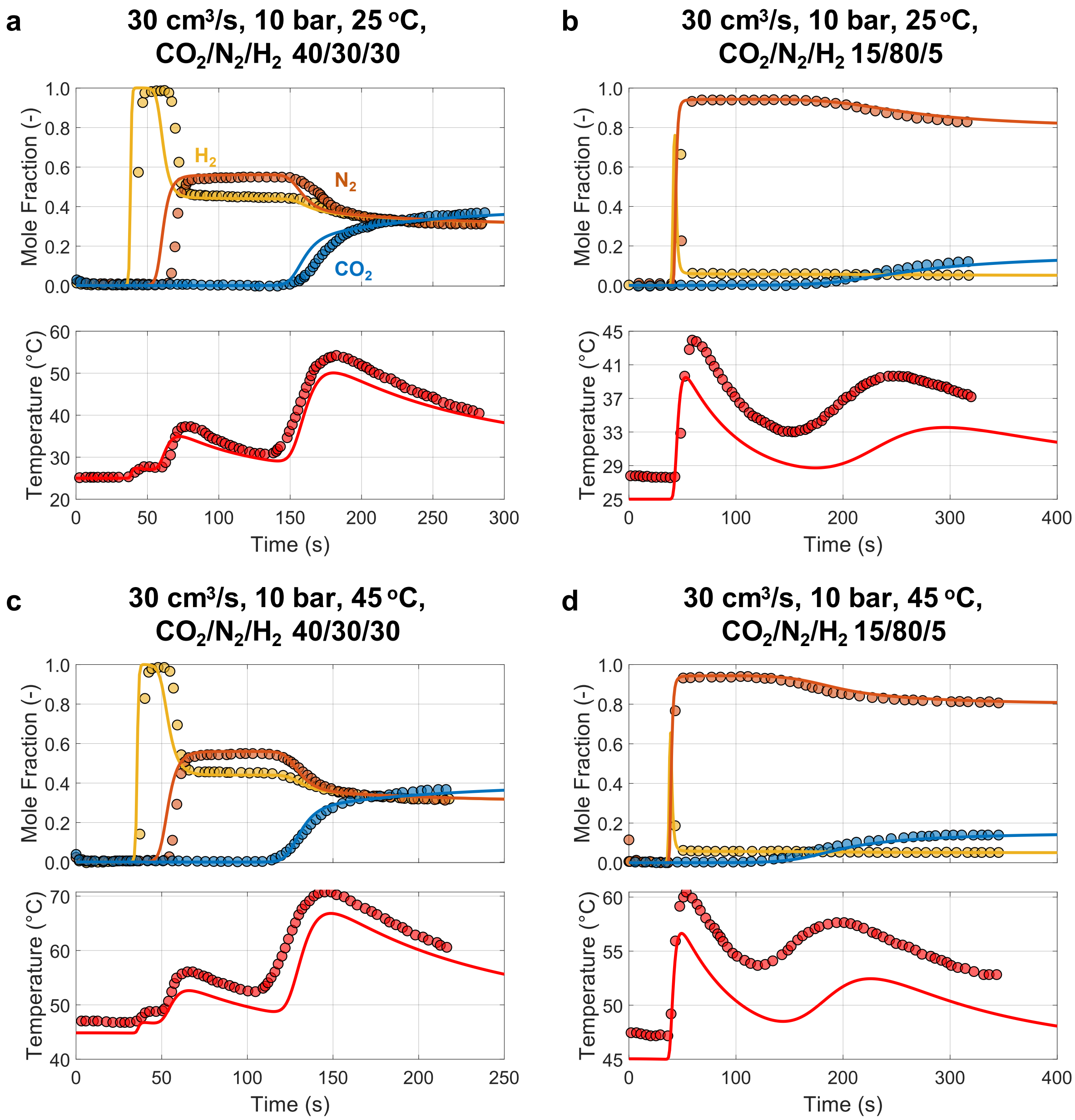

**Fig. 17.** Composition and Temperature breakthrough curves from experiments and BreakLab. Filled circles and solid lines represent the experimental data from Marx *et al.* [70] and BreakLab results, respectively

## 9. Conclusions

AIM is a MATLAB-based GUI workflow software designed to streamline the analysis of adsorbent materials through an integrated computational workflow encompassing isotherm fitting, isosteric enthalpy of adsorption estimation, mixture isotherm prediction, and multicomponent breakthrough simulations. The four modules – IsoFit, HeatFit, MixPred, and BreakLab – offer a user-friendly GUI that supports a wide range of isotherm models (e.g., Langmuir, BET, Toth) and advanced simulation capabilities, such as non-isothermal breakthrough analysis with axial dispersion and Linear Driving Force (LDF) mass transfer. Case studies including Xe/Kr separation in SBMOF-1, $CO_2/CH_4$ separation in CALF-20 demonstrate AIM's functionality and versatility, with results closely matching existing tools such as RUPTURA. Comparison of AIM BreakLab model with experimental data for non-isothermal $CO_2/N_2/H_2$ adsorption in activated carbon establishes the validity of BreakLab mathematical model under different experimental conditions. By providing open-source access to the GUI via GitHub, we aim to promote reproducibility and accessibility in adsorption science and engineering for both experimentalists and computational researchers. However, limitations such as the reliance of constant mass transfer coefficients in BreakLab and challenges in fitting higher-order Virial parameters highlight areas for future refinement and improvement. Future work will focus on incorporating dynamic mass transfer models, expanding available isotherm libraries, and integrating physical properties libraries of different gas components.

**Author Information**

**Corresponding Author**

Yongchul G. Chung – *School of Chemical Engineering, and Graduate School of Data Science, Pusan National University, Busan 46241, Republic of Korea.*

Email: drygchung@gmail.com

**Authors**

Muhammad Hassan – *School of Chemical Engineering, Pusan National University, Busan 46241, Republic of Korea.*

Sunghyun Yoon – *School of Chemical Engineering, Pusan National University, Busan 46241, Republic of Korea.*

Yu Chen – *School of Chemical Engineering, Pusan National University, Busan 46241, Republic of Korea.*

Pilseok Kim – *Department of Chemical and Biological Engineering, Korea University, 145 Anam-ro, Seongbuk-gu, Seoul 02841, Republic of Korea*

Hongryeol Yun – *Department of Chemistry, Korea University, Seoul 02841, Republic of Korea*

Youn-Sang Bae – *Department of Chemical and Biomolecular Engineering, Yonsei University, 50 Yonsei-ro, Seodaemun-gu, Seoul, 03722, Republic of Korea.*

Chung-Yul Yoo – *Department of Energy Systems Research and Chemistry, Ajou University, 206, World Cup-ro, Yeongtong-gu, Suwon 16499, Republic of Korea.*

Dong-Yeun Koh – *Department of Chemical and Biomolecular Engineering (BK-21 Plus), Korea Advanced Institute of Science and Technology (KAIST), Daejeon 34141, Republic of Korea.*

Chang-Seop Hong – *Department of Chemistry, Korea University, Seoul 02841, Republic of Korea*

Ki-Bong Lee – *Department of Chemical and Biological Engineering, Korea University, 145 Anam-ro, Seongbuk-gu, Seoul 02841, Republic of Korea*

**Author Contributions**

**Muhammad Hassan:** Conceptualization, Data curation, Methodology, Formal analysis, Software, Visualization, Writing – original draft, Writing – review and editing. **Sunghyun Yoon:** Methodology, Data curation, Visualization, Writing – review and editing. **Yu Chen:** Data curation. **Pilseok Kim:** Validation, Writing – review. **Hongryeol Yun:** Validation, Writing – review. **Hyuk Take Kwon:** Validation, Writing – review. **Youn-Sang Bae:** Validation, Writing – review. **Chung-Yul Yoo:** Validation, Writing – review. **Dong-Yeun Koh:** Validation, Writing – review. **Chang-Seop Hong:** Validation, Writing – review. **Ki-Bong Lee:** Validation, Writing – review. **Yongchul G. Chung:** Conceptualization, Supervision, Resources, Project administration, Funding acquisition, Writing – review and editing.

**Acknowledgements**

This work was supported by the National Research Foundation of Korea (NRF) from a grant funded by the Korea government (MSIT) (RS-2024-00449431).

**Declaration of Competing Interest**

The authors declare that they have no conflict of interest.

**Data Availability**

The data, MATLAB code, and BreakLab parameters JSON files to reproduce the figures in the manuscript can be found at AIM GitHub repository (https://github.com/Chung-Research-Group/AIM).

**Appendix A. Molecular simulation**

The grand canonical Monte Carlo (GCMC) simulations were conducted to predict the single component adsorption data for $CO_2$/$CH_4$ in CALF-20 and Xe/Kr in SBMOF-1 with a range of pressure (1 Pa up to 1,000,000 Pa) at 298K. Each GCMC simulation consisted of 20,000 cycles, where the initial 10,000 cycles were used for initialization, and the remaining 10,000 cycles were used for the ensemble averages. The swap (insertion and deletion) translation, rotation, and reinsertion Monte Carlo (MC) move were applied for MC sampling with equal probabilities.

The Lennard-Jones 12-6 potential was used to model the non-bonded interaction between zeolites and gases. The TraPPE forcefield [73] was used for $CO_2$/$CH_4$, and the UFF [74]was used to model the Xe/Kr molecule. Dreiding force field [75] was employed for the frameworks. The Lorentz-Berthelot mixing rules were used to calculate the LJ parameters for different atom-type interaction. The van der Waals interactions among atoms in the system were truncated at 14.0 Å with an analytic tail correction. All MC and GCMC simulations were carried out using RASPA 2.0 [76].

## Appendix B. Bayesian inference for uncertainty quantification

The parameters obtained from fitting adsorption isotherm data inherently carry a degree of uncertainty. This uncertainty arises from both measurement uncertainties in the isotherm data and potential limitations of the isotherm model itself. Given that isotherm models are widely used for predicting mixture adsorption and simulating breakthrough behavior, any uncertainty in the model parameters directly impacts the reliability and accuracy of these predictions. Therefore, quantifying the uncertainties associated with isotherm model parameters is essential to ensure the robustness of model predictions. In this study, we employed the uncertainty quantification framework proposed by Ward *et al*. [69], utilizing Bayesian inference analysis for isotherm fitting. Bayesian inference analysis is a well-known tool for the quantification of uncertainties in model parameters given the measured data. In Bayesian inference the model parameters are represented as probability distributions. The mean value of these parameter probability distributions corresponds to the point estimate of the parameter value which would be obtained from traditional model fitting. Additionally, the width of such probability distributions represents the degree of uncertainty in the estimated parameter value. Bayesian inference analysis updates our existing knowledge about model parameters by incorporating the new information obtained from the measured data.

The Bayesian update is based on the Baye's theorem:

$$P(\boldsymbol{\theta}|\mathcal{D}) = \frac{P(\mathcal{D}|\boldsymbol{\theta})\, P(\boldsymbol{\theta})}{P(\mathcal{D})} \tag{B.1}$$

where $\boldsymbol{\theta} = (\theta_1, \theta_2, \theta_3, \ldots, \theta_N)$ are the model parameters, $P(\boldsymbol{\theta})$ is the existing or prior distribution of the model parameters, $\mathcal{D}$ is the set of measured data, $P(\mathcal{D}|\boldsymbol{\theta})$ is the likelihood function and represent the probability of data given the model parameters. $P(\boldsymbol{\theta}|\mathcal{D})$ is the probability of model parameters given the measured data also called posterior distribution. The posterior distribution $P(\boldsymbol{\theta}|\mathcal{D})$ represents how likely the model parameters are given the measured data. The term $P(\mathcal{D})$ is the normalizing constant which ensures that the cumulative area under the posterior distribution is unity. A detailed description of calculating the likelihood function, normalizing constant and prior probability distribution has been provided by Ward *et al.* [69]. Here we only present the resulting expressions.

For the expression of the likelihood function, the errors between the measured data and model predictions for the given set of parameters are assumed to be normally distributed. This assumption yields the following expression for likelihood function:

$$P(\mathcal{D}|\boldsymbol{\theta}) = \frac{1}{\left(\sqrt{2\pi}\,\sigma\right)^{N_p}} \exp\left[-\frac{\sum_{i=1}^{N_p}\left(Q_i^{exp} - Q_i^{sim}\right)}{2\sigma^2}\right] \tag{B.2}$$

where $N_p$ is the number of measurements in the data $\mathcal{D}$. The terms $Q_i^{exp}$ and $Q_i^{sim}$ are the values of the $i^{th}$ measurement and the corresponding model prediction for the given model parameters $\boldsymbol{\theta}$, respectively. The term $\sigma$ is the standard deviation of the error between the data and the model. From the definition of likelihood function, it follows that the $P(\mathcal{D}|\boldsymbol{\theta})$ is high for the model parameters which results in smaller values of residual error.

The expression for the prior probability distribution is derived from the maximum entropy principle and is given as,

$$P(\theta_j) = \frac{1}{\theta_{j,nom}} \exp\left[-\frac{\theta_j}{\theta_{j,nom}}\right] \tag{B.3}$$

where $\theta_{j,norm}$ is the nominal or fitted value of the $j^{th}$ model parameter $\theta_j$. Equation (B.3) is the marginal prior distribution depending only on the given parameter $\theta_j$. Assuming that all the model parameters are independent of one another, a joint prior probability distribution considering all the model parameters $\boldsymbol{\theta} = (\theta_1, \theta_2, \theta_3, \dots, \theta_N)$ can be obtained as follows,

$$P(\boldsymbol{\theta}) = \prod_{j=1}^{n_p}\left(\frac{1}{\theta_{j,nom}} \exp\left[-\frac{\theta_j}{\theta_{j,nom}}\right]\right) \tag{B.4}$$

where $n_p$ is the total number of model parameters.

The normalizing constant $P(\mathcal{D})$ is obtained by integrating the joint probability distributions of the data given the model parameters $P(\mathcal{D}|\boldsymbol{\theta})$ and prior probability of model parameters $P(\boldsymbol{\theta})$ over the full domain of model parameters.

$$P(\mathcal{D}) = \int_{\theta} P(\mathcal{D}|\boldsymbol{\theta})\, P(\boldsymbol{\theta}) d\boldsymbol{\theta} \tag{B.5}$$

Using equations (B.2), (B.4), and (B.5), we can determine the joint posterior distribution $P(\boldsymbol{\theta}|\mathcal{D})$. The marginal probability distribution for each of the model parameters can be obtained by marginalizing equation (B1) that is,

$$P(\theta_j|\mathcal{D}) = \int_{\theta_{-j}} \frac{P(\mathcal{D}|\boldsymbol{\theta})\, P(\boldsymbol{\theta})}{P(\mathcal{D})} d\boldsymbol{\theta}_{-\boldsymbol{j}} \tag{B.6}$$

where $\theta_{-j}$ represents that integral is taken with respect to all the model parameters except $\theta_j$.

Together equations (B.2-B.6) constitute Bayesian inference for obtaining posterior probability distribution of model parameters. The integrals in equations (B.5) and (B.6) are multidimensional and are evaluated using Monte Carlo integration techniques. A MATLAB based implementation of Bayesian inference using quasi-Monte Carlo integration has been provided by Ward *et al.* [69]. We used the supplied code and coupled it with our isotherm functions to obtain uncertainty in the model parameters for the case studies.